\documentclass{article} 
\usepackage{iclr2027_conference,times}

\usepackage{amsmath,amsfonts,bm}

\def\eqref#1{equation~\ref{#1}}

\def\1{\bm{1}}

\DeclareMathAlphabet{\mathsfit}{\encodingdefault}{\sfdefault}{m}{sl}
\SetMathAlphabet{\mathsfit}{bold}{\encodingdefault}{\sfdefault}{bx}{n}

\usepackage{hyperref}
\usepackage{url}
\usepackage{dashrule}
\newcommand{\partitle}[1]{\smallskip \noindent \textbf{#1.}}
\newcommand{\name}{{\tt AuxMark}}

\title{AuxMark: Defending Against Unauthorized Agent Distillation via Auxiliary Behavioral Watermarking}

\author{
Yiqing Feng\textsuperscript{1},
Haozhe Feng\textsuperscript{2},
Shunan Shang\textsuperscript{1},
Xiaoyu Zhang\textsuperscript{1},
Jian Lou\textsuperscript{3},\\
\ \textbf{Haodong Zhao\textsuperscript{4},
Mingxun Zhou\textsuperscript{5}}\thanks{Corresponding author.}\\
\textsuperscript{1}Xidian University,
\textsuperscript{2}Zhejiang University\\
\textsuperscript{3}Sun Yat-sen University,
\textsuperscript{4}Shanghai Jiao Tong University,
\textsuperscript{5}HKUST
}
\newcommand{\tableRate}[3]{#1 / #2 (#3\%)}
\newcommand{\tablePval}[2]{\ensuremath{#1 \times 10^{#2}}}
\definecolor{colorLLM4}{HTML}{EAF3E2}
\definecolor{colorLLM3}{HTML}{B4DEB6}
\definecolor{colorLLM2}{HTML}{7BC6BE}
\definecolor{colorLLM1}{HTML}{439CC4}
\newcommand{\robustnessModelLegend}{bar colors denote \textcolor{colorLLM4}{$\blacksquare$} Mistral-24B, \textcolor{colorLLM3}{$\blacksquare$} GLM-4.7-Flash, \textcolor{colorLLM2}{$\blacksquare$} Qwen3-14B, and \textcolor{colorLLM1}{$\blacksquare$} Qwen3-32B}
\newcommand{\robustnessPanelWidth}{0.31\textwidth}
\usepackage{algorithm}
\usepackage{algpseudocode} 

\usepackage{amsmath}
\usepackage{amssymb}
\usepackage{graphicx}
\usepackage{subcaption}
\usepackage{multirow}
\usepackage{wrapfig}
\usepackage{placeins}

\iclrfinalcopy 
\begin{document}

\maketitle

\begin{abstract}
Large language model agents can acquire complex capabilities through multi-step interaction and tool use, but their trajectories can also be illegally collected to distill student agents. However, existing watermarking methods either do not fit the structured and interactive nature of agent environments or lack reliable effectiveness across tasks and model architectures. We introduce \name{}, a behavioral watermarking framework for tracing unauthorized agent distillation. \name{} dynamically inserts safe, non-essential auxiliary action into teacher trajectories, and stores the associated contexts as private evidence cards. To audit a suspicious student model, \name{} constructs paired real and fake probes from these cards and applies a card-level sign test. This black-box protocol supports both model-level detection and trace-level attribution. Across three agent benchmarks, two teacher agents, and four student architectures, \name{} detects all 24 distilled models with zero false positives on 48 clean models. It also preserves task utility and remains effective against data flooding, paraphrasing, truncation, and adaptive cleaning attacks. Our code will be released at \href{https://github.com/qx041609/Auxmark}{this URL}.
\end{abstract}
\vspace{-8pt}
\section{Introduction}
\vspace{-8pt}
\label{sec:introduction}
Large language model (LLM) agents can solve complex tasks by interacting with external environments and using tools~\citep{anthropic2026opus48, team2025kimi, xu2026deepseek}. However, training models with these advanced skills requires huge costs and computing resources. For example, training frontier models like OpenAI's GPT requires tens of thousands of GPUs and costs tens of millions of dollars~\cite{singh2025openai, sajadieh2026artificial}. Agent distillation is a cheap way to significantly improve model performance~\cite{liu2026structured, kang2026distilling}, so it has become widely used. Competitors systematically query a powerful teacher agent to collect high-quality interaction trajectories, using them to fine-tune their own student models to avoid the high costs of training from scratch. This unauthorized knowledge transfer severely harms the intellectual property of model developers. This threat is not just a theory; leading AI companies (e.g., OpenAI, Google) have already found massive illegal distillation behaviors in their traffic~\cite{OpenAIkey, googleGTIGThreat, anthropicDetectingPreventing}. Furthermore, recent industry disputes and studies have exposed widespread trajectory copying, showing that many models display ``behavioral homogenization" that closely mimics the proprietary teachers~\cite{yang2026agents}. 

To protect model intellectual property, several methods have been proposed, but they face different critical limitations when applied to agent distillation. First, antidistillation watermarks designed for language models modify text tokens or reasoning traces~\cite{savani2026antidistillation, xu2026antidistillation, ma2026protecting}, which breaks the strict format required for agent tool calls. Second, existing agent antidistillation watermarking schemes usually assume that the teacher and student share the same base model, resulting in poor performance during cross-architecture distillation~\cite{wang2026protecting}. Third, unwatermarked fingerprinting methods based on execution similarity~\cite{yang2026agents, rawat2026reference} are unreliable, because high behavioral similarity often comes from the standard solutions of the tasks rather than actual distillation. Therefore, finding a reliable way to trace and prove this unauthorized distillation through black-box interactions has become an urgent challenge.

In this work, we propose \name{}, a behavioral watermarking framework to trace agent distillation. Our main insight stems from the multi-step nature of agent interactions and the behavioral homogenization during distillation. Instead of modifying the core actions required for the task, \name{} injects extra auxiliary actions into the trajectory and saves these interaction as private evidence. To align with deployments in the real world, we introduce a more practical threat model: model owners can identify suspicious distillation behaviors via traffic monitoring (existing studies show that such traffic anomaly detection can achieve around 90\% recall with extremely low false positive rates~\cite{liu2026embarrassingly, huh2025preventing, chiang2024chatbot}), and extract the corresponding evidence for targeted verification. For this verification, we introduce a paired-probe detection protocol. It compares the suspect model's responses on real probes built from the evidence against fake probes where the core actions are semantically altered. This strict comparison effectively isolates the watermarked behavior from natural contextual choices. Empirically, our evaluation demonstrates that \name{} achieves highly reliable detection while preserving the agent's task utility. Across three agent benchmarks, two teacher models, and four student architectures, our method successfully detects all 24 distilled student models with zero false positives. Furthermore, \name{} remains highly robust against various adversarial attacks.

\textbf{Key contributions of this paper include:}
\vspace{-5pt}
\begin{itemize}
    \item We formulate a highly practical threat model for agent distillation. By combining suspicious target identification via traffic monitoring with targeted verification using private evidence, this model provides a closed-loop solution for tracing intellectual property theft in the real world scenarios.
    \item We design a novel behavioral watermarking framework and a paired-probe detection protocol. This architecture exploits multi-step interactions to embed auxiliary actions in real time, and provides strong statistical guarantees for verification and trace-level attribution by comparing responses on real and fake probes alongside a card-level sign test.
    \item We conduct comprehensive experiments across multiple benchmarks and model architectures. The results demonstrate that \name{} achieves a 100\% detection rate and zero false positives while preserving the agent's task utility, and exhibits strong robustness against various data-processing operations and adaptive cleaning attacks.
\end{itemize}
\vspace{-8pt}
\section{Problem Formulation and Preliminaries}
\label{sec:preliminaries}
\vspace{-8pt}
\subsection{Agent Distillation}
\vspace{-5pt}
LLM agent interacts with an external environment to solve a user task through multiple rounds of reasoning and tool use~\cite{yao2022react, qin2024toolllm}. An agent trajectory records intermediate reasoning and tool-use behaviors. Given a user query $q$, an LLM agent $\mathcal{M}$ interacts with the environment over $T$ steps. At each step $i$, based on the query $q$ and the interaction history $h_i$, the agent generates a thought $t_i$ and an action $a_i=(f_i,p_i)$, where $f_i$ denotes the selected tool and $p_i$ denotes its parameters. The environment executes the action and returns an observation $o_i$. A complete interaction trajectory $\tau$ is formally represented as:
\begin{equation}
    \tau = \{(t_i,a_i,o_i)\}_{i=1}^{T}, \text{ where } (t_i, a_i) = \mathcal{M}(q, h_i).
    \label{equ:trajectory}
\end{equation}
Agent distillation \cite{kang2026distilling,luo2026agentark,liu2026structured} aims to transfer such interaction behaviors from a powerful teacher agent $\mathcal{M}_t$ to a student agent $\mathcal{M}_s$. The teacher first generates a set of $N$ trajectories $(q_j,\tau_j)_{j=1}^{N}$. In agent distillation, we consider hard distillation based on supervised fine-tuning \cite{ouyang2022training}. Through this process, the student can learn the teacher's reasoning and tool-use behaviors from its interaction trajectories. The student model is trained on $(q_j,\tau_j)_{j=1}^{N}$, where previous thoughts, actions, and observations are used as context, while the teacher's thoughts and actions are used as supervision~\cite{kang2026distilling,luo2026agentark}.
\vspace{-5pt}
\subsection{Model Watermarking}
\vspace{-5pt}
To protect the intellectual property of a teacher agent $\mathcal{M}_t$, the model owner employs a secret key $k$ to embed watermark signals in real time during the generation of trajectories $\tau$. After attackers query the model to obtain a watermarked dataset $\mathcal{D}$, they attempts to clone the teacher's capabilities by training an base model $\mathcal{M}_c$ on $\mathcal{D}$ via a distillation algorithm to get the student model $\mathcal{M}_s$. Then model owner audits suspicious student models using a private probe set $\mathcal{P}_k$. A practical anti-distillation watermark should satisfy the following three  requirements:

\textbf{Effectiveness.} An effective watermark must reliably transfer to the student model during distillation while maintaining a minimal false positive rate on independent models. During verification, the owner queries the suspicious model using the probe set $\mathcal{P}_k$ and computes a $p$-value to evaluate the watermark signal. A distilled student model $\mathcal{M}_s$ should exhibit statistically significant evidence of the watermark, satisfying $p\mathtt{-value} < p$, whereas a clean model $\mathcal{M}_{c}$ should yield $p\mathtt{-value} > p$.

\begin{wrapfigure}[17]{r}{0.5\columnwidth}
    \vspace{-\intextsep}
    \centering
    \includegraphics[width=\linewidth]{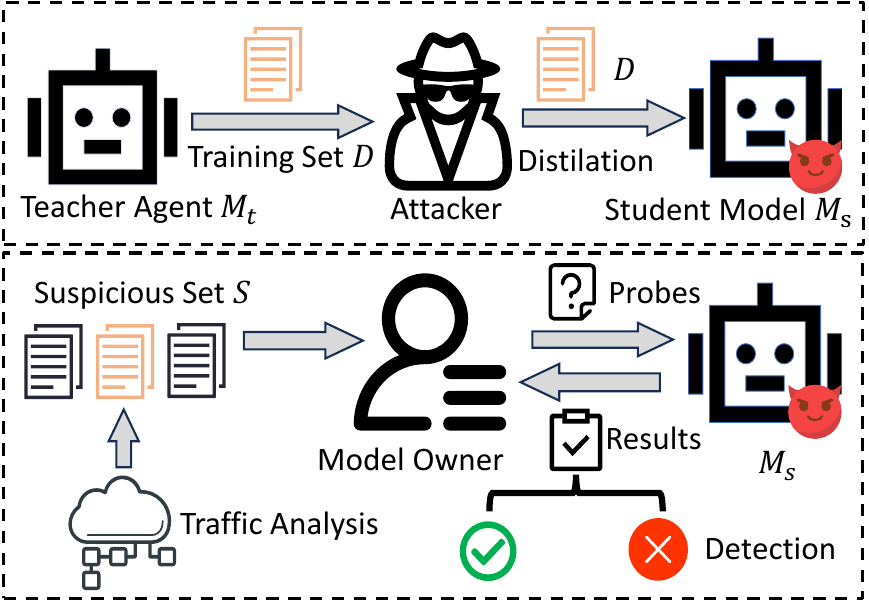}
    \vspace{-15pt}
    \caption{Overview of the access and threat model. The attacker distills a student model from trajectories, while the model owner identifies suspicious traffic and audits the student.}
    \label{fig:threat-model}
\end{wrapfigure}
\textbf{Harmlessness.} The real-time watermarking mechanism should not degrade the normal operation of the teacher agent $\mathcal{M}_t$. This entails two main aspects: task utility and inference efficiency. Specifically, the watermarked teacher model must maintain a success rate on the same tasks comparable to its unwatermarked counterpart, and the watermark embedding process should not significantly inflate the length or computational cost of the generated trajectories.

\textbf{Robustness.} The watermark must resist data modifications applied by attackers. Even if the attacker alters the dataset through data flooding, semantic rewriting, truncation, and adaptive cleaning attacks. the watermark signal should remain detectable in the student model.
\vspace{-5pt}
\subsection{Access and Threat Model}
\vspace{-5pt}
We formulate a two-party threat model comprising a model owner and an attacker as shown in Figure~\ref{fig:threat-model}. The owner deploys a teacher agent $\mathcal{M}_t$, while the attacker aims to distill a student model $\mathcal{M}_s$ using a training dataset $\mathcal{D}$ collected from the teacher agent.

\partitle{Model Owner} The owner deploys and fully controls a teacher agent $\mathcal{M}_t$ (e.g., Codex~\cite{singh2025openai}) as a service. To protect the model, the owner embeds watermarks into the output trajectories in real time. Since data scraping has specific patterns, the owner monitors server-side features~\cite{huh2025preventing, chiang2024chatbot, liu2026embarrassingly}. Any interactions showing such signs are recorded to build a suspicious trajectory set $\mathcal{S}$. To avoid missing potential distillation traces, the owner conservatively records all suspicious traces into $\mathcal{S}$ and the owner cannot determine which traces in $\mathcal{S}$ were actually used for distillation. When testing the deployed student model $\mathcal{M}_s$, the owner only has black box access and can fully control the API inputs to verify the watermark.

\partitle{Attacker}
To bypass the high training costs, the attacker aims to clone the capabilities of the teacher $\mathcal{M}_t$. Operating under strict black-box access, the attacker systematically queries the teacher agent $\mathcal{M}_t$ to harvest complete multi-step interaction trajectories and this large-scale automated querying leaves identifiable server-side traffic patterns that can be captured by the model owner. The attacker may then preprocess these collected trajectories through operations such as sanitization, rewriting, truncation, or mixing with other data, forming the final distillation dataset $\mathcal{D}$. Finally, the attacker uses $\mathcal{D}$ to distill the student model $\mathcal{M}_s$ and deploys it as a black-box API.
\vspace{-8pt}

\section{Watermark Construction}
\vspace{-10pt}
\subsection{Overview}
\begin{wrapfigure}[23]{r}{0.48\textwidth}
\vspace{-\intextsep}
\centering
\includegraphics[width=\linewidth]{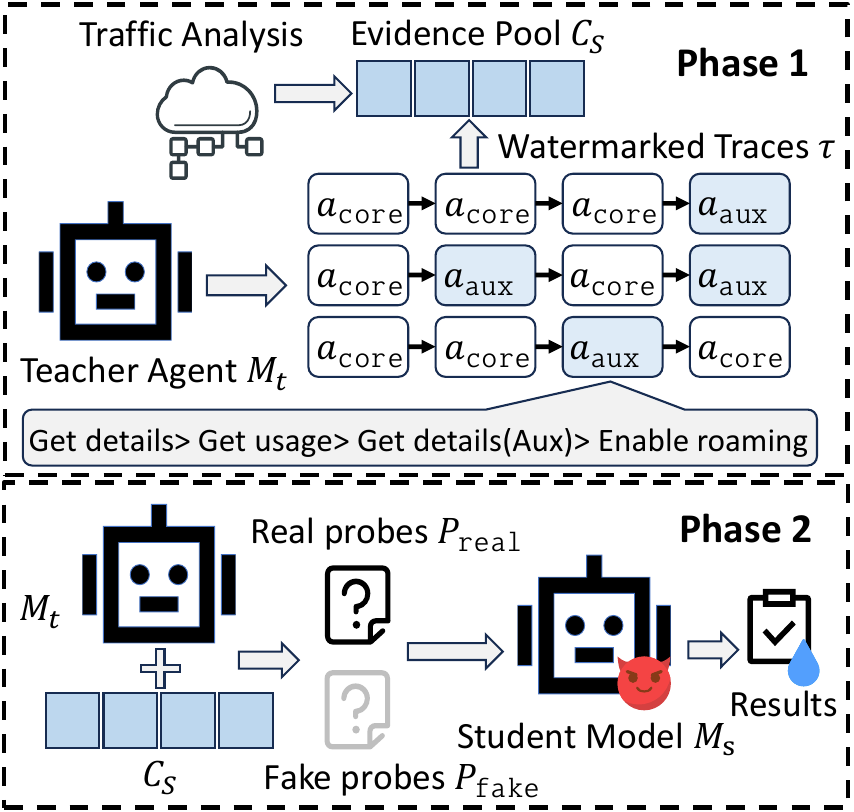}
\vspace{-15pt}
\caption{Overview of \name{}. \textbf{Phase 1:} the owner embeds the watermark by inserting aux behaviors; \textbf{Phase 2:} the owner retrieves the private evidence to construct paired real and fake probes and verify a suspect student model. \textcolor[HTML]{BDD7EE}{\rule{0.85em}{0.85em}} denotes watermarked thing.}
\label{fig:overview}
\end{wrapfigure}

\vspace{-5pt}
During agent distillation, the distilled students and their teachers share similar interaction trajectories~\cite{yang2026agents, lyu2025correction, gudibande2023false}. Based on this property, we proposed a watermark scheme \name{} as illustrated in Figure~\ref{fig:overview}. Firstly, the owner embeds the watermark by dynamically inserting auxiliary(aux) behaviors into the teacher's trajectories and privately saves evidence. Once suspicious traffic is identified, the corresponding records are extracted to form an evidence pool in Section~\ref{sec:watermark_embedding}. Next, the owner uses this evidence to construct paired real and fake probes, applying statistical tests to verify if the suspect model reproduces the watermark in Section~\ref{Detection}. Finally, trace-level attribution in Section~\ref{subsec:trace_level} to evaluate which suspicious trajectories are most likely to have been used in the attacker's training set.

\subsection{Watermark Embedding Algorithm}
\vspace{-5pt}
\label{sec:watermark_embedding}
Unlike previous watermarking methods that directly change the generated text or the teacher model's actions~\cite{wang2026protecting,an2026sequential}, \name{} embeds watermarks by adding extra, non-essential auxiliary actions into the agent's trajectory. By doing this, \name{} keeps the teacher model's performance high while creating strong evidence for later tracing. Also, to stay hidden, the added aux actions look and act exactly like normal tool calls: they use valid tools and bind strictly to normal parameter values derived from the interaction history. Based on the threat model described in Section \ref{sec:preliminaries}, the owner can subsequently identify suspicious traces from the API traffic and export their privately saved watermark evidence for later verification. The entire embedding process consists of three main stages shown in Figure\ref{fig:watermark_embedding_protocol}: safe tool classification, dynamic scheduling and candidate generation, and private evidence retention.

\partitle{Safe tool classification} To ensure aux actions will not change the environment state, \name{} applies a conservative safety policy to the tool schema $\mathcal{F}$ of trace $\tau$. Since the system operates as an agent-as-a-service~\cite{li2026aidev}, this safety evaluation is performed in advance and cached to avoid real-time latency. An offline safety model $\mathcal{M}_{\mathtt{safe}}$ evaluates the tool schema $\mathcal{F}$ and maps each tool $f \in \mathcal{F}$ to a safety label as shown in Eq.\ref{equ:safeclassfier}:
\begin{equation}
\vspace{-2pt}
\mathcal{M}_{\mathtt{safe}}(f)=\begin{cases}\mathtt{safe},&\text{if }f\text{ is read-only,}\\\mathtt{unsafe},&\text{otherwise.}\end{cases}
\vspace{-2pt}
\label{equ:safeclassfier}
\end{equation}
The system exclusively permits tools from the valid subset $\mathcal{F}_{\mathtt{safe}}=\{f \in \mathcal{F} \mid \mathcal{M}_{\mathtt{safe}}(f)=\mathtt{safe}\}$, categorically rejecting any action that changes the environment state or influences the task process.

\partitle{Dynamic scheduling} Having established the safe tool subset $\mathcal{F}_{\mathtt{safe}}$, the system must determine when to invoke these aux action during the interaction as the watermarking. To make watermark positions unpredictable, \name{} dynamically evaluates whether to inject an aux action at step $t$ immediately after the core action $a_{\mathtt{core}}^{t-1}$ finishes. If skipped, the system waits and checks again after the next core action. For each trajectory $\tau$ named as $\mathtt{id}_\tau$ of an account $a$, the system keeps a release probability $p_t$ and a remaining budget $b_t$ at step $t$. Immediately after the core action $a_{\mathtt{core}}^{t-1}$ finishes, the system uses a cryptographic hash function $\mathtt{Hash}(\cdot) \to [0,1)$~\cite{preneel1994cryptographic} parameterized by the owner's secret key $k$ to compute a uniformly distributed random value $u_t$, as defined in Eq.~\ref{equ:hash}:
\begin{equation}
    \vspace{-2pt}
    \label{equ:hash}
u_t = \operatorname{\mathtt{Hash}}(k \mathbin\Vert a \mathbin\Vert \mathtt{id}_\tau \mathbin\Vert t) \in [0, 1).
\vspace{-2pt}
\end{equation}
The system tries to add an aux action when $b_t>0$ and $u_t<p_t$. Depending on whether the auxiliary action is successfully inserted or the turn is skipped (due to generation failure or unmet conditions), the system updates the trigger probability and remaining budget for the next turn as follows:
\begin{equation}
    \vspace{-2pt}
\label{equ:pb_update}
(p_{t+1}, b_{t+1}) = 
\begin{cases}
(p_0, b_t - 1), & \text{if successfully inserted,} \\
(\min\{p_t + \Delta p, p_{\mathtt{max}}\}, b_t), & \text{otherwise.}
\end{cases}
\vspace{-2pt}
\end{equation}
\partitle{Aux action embedding} If the dynamic scheduling condition is met, \name{} proceeds to generate and select the optimal aux action. To obtain high quality insertions, the system queries $\mathcal{M}_t$ to propose up to $K$ candidates. Each candidate $j$ is formulated as a tuple of a thought and an action: $(t_{\mathtt{aux}}^j, a_{\mathtt{aux}}^j)$, where the action $a_{\mathtt{aux}}^j$ consists of a tool $f_{\mathtt{aux}}^j$ and its arguments $p_{\mathtt{aux}}^j$. To ensure that the injected action perfectly blends into the trajectory, \name{} enforces a validation mechanism. First, the chosen tool must belong to the safe subset $\mathcal{F}_{\mathtt{safe}}$. Second, to prevent model hallucinations and ensure contextual rationality, the arguments $p_{\mathtt{aux}}^j$ must be extracted from the context (such as the core action's observation or history). Consequently, \name{} categorically discards any candidate that uses an unapproved tool or hallucinates untraceable arguments.

For each candidate $j$ that passes the validation, \name{} employs a scoring model $\mathcal{M}_q$ to evaluate it across three dimensions: naturalness, logical consistency, and distillability. Specifically, $\mathcal{M}_q$ assigns a relevance score $s_{\mathtt{rel}}^j$ to ensure the injected action aligns naturally with the task. Additionally, a reliability score $s_{\mathtt{relb}}^j$ ensures the high quality of the thought. Crucially, to foster a stable behavioral habit that the student model can easily absorb during distillation, the system calculates a repeatability score $s_{\mathtt{rep}}^j$. This score is directly determined by $n_j$, which represents the number of times the candidate's specific tool pairing $(f_{\mathtt{core}}, f_{\mathtt{aux}}^j)$ has been successfully injected in the account's history. The final score $S_j$ for candidate $j$ is computed as a weighted sum:
\begin{equation}
    \vspace{-4pt}
    S_j = \alpha \cdot s_{\mathtt{rel}}^j + \beta \cdot s_{\mathtt{relb}}^j + \gamma \cdot s_{\mathtt{rep}}^j.
    \label{equ:score}
    \vspace{-2pt}
\end{equation}
where $\alpha$, $\beta$, $\gamma$, and the step-function mapping from repeat count $n_j$ to $s_{\mathtt{rep}}^j$ are detailed in Appendix~\ref{sec:experimental-details}. The system selects the candidate with the highest score as the best aux action $(t_{\mathtt{aux}}^{\mathtt{best}}, a_{\mathtt{aux}}^{\mathtt{best}})$, then injects this aux action, establishing a natural sequence:
\begin{equation}
    \vspace{-2pt}
    t_{\mathtt{core}}^{t-1} \;\rightarrow\; a_{\mathtt{core}}^{t-1} \;\rightarrow\; o_{\mathtt{core}}^{t-1} \;\rightarrow\; t_{\mathtt{aux}}^{\mathtt{best}} \;\rightarrow\; a_{\mathtt{aux}}^{\mathtt{best}} \;\rightarrow\; o_{\mathtt{aux}}^{\mathtt{best}}.
    \vspace{-2pt}
\end{equation}
Based on the generation result, the system updates its states. A successful generation applies the success update (Eq.~\ref{equ:pb_update}) and increases the repeat count $n_{\mathtt{best}}$ for this tool pairing. If the generation fails, the system skips the insertion, applies the failure update, moves to generate core action and waits for next chance. For future auditing, every successful insertion creates a private evidence card $c_t$. To securely record the exact context and execution details, this card stores the historical context $h$, the preceding core action and observation, and the injected auxiliary tool parameters. It is saved as a tuple and added to the trajectory's evidence set $\mathcal{C}_\tau$:
\begin{equation}
    \vspace{-2pt}
    c_t = (h, t_{\mathtt{core}}^{t-1}, a_{\mathtt{core}}^{t-1}, o_{\mathtt{core}}^{t-1}, f_{\mathtt{aux}}^{\mathtt{best}}, p_{\mathtt{aux}}^{\mathtt{best}}), \quad \mathcal{C}_\tau = \mathcal{C}_\tau \cup \{c_t\}.
    \vspace{-2pt}
    \label{equ:evidence}
\end{equation}
As mentioned in our threat model, the model owner uses signals like querying styles, IPs, and traffic patterns to identify a suspicious set $\mathcal{S}$~\cite{chiang2024chatbot, liu2026embarrassingly}. The owner then collects the evidence from these flagged traces to form a global verification pool $\mathcal{C}_{\mathcal{S}} = \bigcup_{\tau \in \mathcal{S}} \mathcal{C}_\tau$.

\vspace{-5pt}
\subsection{Watermark Detection}
\vspace{-5pt}
\label{Detection}
While the embedding phase injects an aux action into the teacher's trajectory, the detection phase verifies if a suspect model $M_s$ has learned this specific behavior pattern. To achieve this, \name{} uses paired probes built from the combined evidence pool $\mathcal{C}_{\mathcal{S}}$ for verification as shown in Figure~\ref{fig:watermark_detection_protocol}.

\partitle{Paired probe construction} To verify watermark retention, \name{} constructs paired probes for the combined evidence pool $\mathcal{C}_{\mathcal{S}}$. A probe is formally defined as a prompt-target tuple: given a history and a core action as the prompt, the suspect model $\mathcal{M}_s$ is expected to output a specific aux action. Specifically, based on each evidence card, we derive two types of probes: real probes and  fake probes. This paired design ensures that the model $\mathcal{M}_s$ has genuinely memorized the specific watermarked trace, rather than unconditionally outputting the target action.

For the real side construction of each card $c_i \in \mathcal{C}_S$, the system first locates the nearest core action. To ensure that the suspect model $M_s$ has learned the expected behavior, \name{} creates up to $N$ parameterized variants for this real side by modifying one argument of the original core action, the system prompts the teacher model $\mathcal{M}_t$ to synchronously generate a logically consistent core triplet $(t_{\mathtt{core}}^j, a_{\mathtt{core}}^j, o_{\mathtt{core}}^j)$ alongside the new downstream target aux arguments $p_{\mathtt{aux}}^j$. This process directly generates a set of real side probes $p_{\mathtt{real}}^i$ for the specific card $c_i$ as shown in Eq.\ref{equ:probe real}:
    \vspace{-2pt}
\begin{equation}
    \label{equ:probe real}
p_{\mathtt{real}}^i = \left\{ (h, t_{\mathtt{core}}^j, a_{\mathtt{core}}^j, o_{\mathtt{core}}^j),\ (f_{\mathtt{aux}}, p_{\mathtt{aux}}^j)  \right\}_{j=1}^{N_i}.
\vspace{-2pt}
\end{equation}
$h$ is the historical context, and $f_{\mathtt{aux}}$ is the targeted aux tool. Finally, the system aggregates the real-side probes from all cards to form the global real probe set $\mathcal{P}_{\mathtt{real}} = \bigcup_{c_i \in \mathcal{C}_S} p_{\mathtt{real}}^i$.

To better contrast with the real probes $\mathcal{P}_{\mathtt{real}}$, obtain more definitive statistical results, \name{} constructs a set of fake probes based on the $\mathcal{P}_{\mathtt{real}}$. This verifies whether the aux action stems from memorization rather than natural context or random chance.For this fake side. However, it prompts the teacher model $\mathcal{M}_t$ to replace the core action with a semantically similar fake core behavior $(\hat{t}_{\mathtt{core}}^j, \hat{a}_{\mathtt{core}}^j,\hat{o}_{\mathtt{core}}^j)$ that achieves the identical goal. This process generates the fake probe set $p_{\mathtt{fake}}^i$ for card $c_i$, directly matching the variants in $p_{\mathtt{real}}^i$ as shown in Eq.\ref{probe fake}:
    \vspace{-2pt}
\begin{equation}
    \label{probe fake}
p_{\mathtt{fake}}^i = \left\{  (h, \hat{t}_{\mathtt{core}}^j, \hat{a}_{\mathtt{core}}^j, \hat{o}_{\mathtt{core}}^j),\ (f_{\mathtt{aux}}, p_{\mathtt{aux}}^j)  \right\}_{j=1}^{N_i}.
\vspace{-2pt}
\end{equation}
Thus, the strict difference between the paired probes lies only in the core action. Similar to the real side, the fake probe set is defined as $\mathcal{P}_{\mathtt{fake}} = \bigcup_{c_i \in \mathcal{C}_S} p_{\mathtt{fake}}^i$.

\partitle{Card-level statistical test} After generating the paired probes, \name{} evaluates the suspect model $M_s$ using $(\mathcal{P}_{\mathtt{real}}, \mathcal{P}_{\mathtt{fake}})$ and performs a card-level statistical test which is discussed in Appendix~\ref{sec:statistical-test-discussion}. For the $j$-th probe pair of card $c_i$, the expected aux action is $f_{\mathtt{aux}}$ and arguments $p_{\mathtt{aux}}^j$. Let $(f_{\mathtt{out}}, p_{\mathtt{out}})$ be the tool call parsed from the output of $M_s$. For both the real and fake sides, \name{} computes the toolhit indicator $t_{i,j}$ and the strict fullhit indicator $h_{i,j} \in \{0, 1\}$:
\vspace{-2pt}
\begin{equation}
t_{i,j} = \operatorname{\mathtt{Match}}(f_{\mathtt{out}}, f_{\mathtt{aux}}), \quad h_{i,j} = t_{i,j} \cdot \operatorname{\mathtt{Match}}(p_{\mathtt{out}}, p_{\mathtt{aux}}^j).
\label{eq:probe-score}
\end{equation}
where the function $\operatorname{\mathtt{Match}}(\cdot, \cdot) \to \{0, 1\}$ evaluates to $1$ if and only if there is exact structural and value equivalence between its two arguments, and $0$ otherwise. Because the $N_i$ paired variants derived from the same card $c_i$ share an identical core-to-aux dependency, they are not independent samples. Therefore, \name{} aggregates the scores at the card level to calculate the real-side hit rate $R_i^{\mathtt{real}}$, the fake-side hit rate $R_i^{\mathtt{fake}}$, and their difference $\Delta_i$ for each card $c_i$ as shown in Eq.\ref{eq:probe-rate}:
\begin{equation}
    \vspace{-2pt}
    R_i^{\mathtt{real}} = \frac{1}{N_i} \sum_{j=1}^{N_i} h_{i,j}^{\mathtt{real}},\qquad
R_i^{\mathtt{fake}} = \frac{1}{N_i} \sum_{j=1}^{N_i} h_{i,j}^{\mathtt{fake}},\qquad
\Delta_i = R_i^{\mathtt{real}} - R_i^{\mathtt{fake}}.
\vspace{-2pt}
\label{eq:probe-rate}
\end{equation}
To evaluate the verification, we establish the null hypothesis $H_0$ and the alternative hypothesis $H_1$. $H_0$ posits that the suspect model $\mathcal{M}_s$ has not learned the watermarked habit; any correct action is due to general context or innate biases. Conversely, $H_1$ asserts that $\mathcal{M}_s$ has memorized the watermark, making it significantly more likely to output the target action on real probes. Based on the performance difference $\Delta_i$ for each card $c_i$, we classify the outcomes into a real win ($\Delta_i > 0$), a fake win ($\Delta_i < 0$), or a tie ($\Delta_i = 0$). Excluding ties, let $w$ and $\ell$ be the numbers of real and fake wins across $\mathcal{C}_{\mathcal{S}}$. The $p$-value is computed using a one-sided exact sign-test~\cite{hollander2013nonparametric}:
\vspace{-2pt}
\begin{equation}
    p\mathtt{-value} = \sum_{k=w}^{w+\ell}\binom{w+\ell}{k}2^{-(w+\ell)}.
    \vspace{-2pt}
\end{equation}
where $p\mathtt{-value} = 1$ if $w+\ell=0$. To reject $H_0$ and declare a successful verification, \name{} requires dual criteria: the statistical significance must be $p\mathtt{-value} < p$, and the averaged real fullhit rate must exceed the fake rate by a predefined margin $r$ (i.e., $\frac{1}{|\mathcal{C}_{\mathcal{S}}|} \sum \Delta_i \geq r$). This empirical margin absorbs unpredictable fluctuations caused by context rewriting in the fake probes and guards against false positives. Appendix~\ref{sec:experimental-details} specifies the hyperparameters $p$, $r$, and $N$.

\vspace{-5pt}
\subsection{Trace-Level Attribution} \label{subsec:trace_level}
\vspace{-5pt}
While the detection protocol in Section~\ref{Detection} determines whether a suspect model $\mathcal{M}_s$ has distilled from $\mathcal{M}_t$, it cannot pinpoint which traces within $\mathcal{S}$ were utilized. For exact attribution, \name{} introduces a trace-level attribution. For each evaluated trace $\tau$, the system computes four rates to capture both exact and partial matches: the strict full-hit rates $R_\tau^{\mathtt{real}}$ and $R_\tau^{\mathtt{fake}}$, and the toolhit-only rates $T_\tau^{\mathtt{real}}$ and $T_\tau^{\mathtt{fake}}$ (where $t_{i,j}=1 \land h_{i,j}=0$). To ensure fair comparison across traces, \name{} normalizes each rate into an average-rank quantile $Q(\cdot) \in (0, 1)$:
\vspace{-2pt}
\begin{equation}
    Q(x_\tau) = \frac{\operatorname{\mathtt{rank}}_{\mathtt{avg}}(x_\tau)}{|\mathcal{S}| + 1}.
\vspace{-2pt}
\end{equation}
The attribution logic is designed as follows: $Q(R_\tau^{\mathtt{real}})$ acts as the primary positive signal for trace memorization. Conversely, if the model outputs the target aux action even on the fake probes, it implies innate biases rather than specific watermark retention. Thus, $Q(R_\tau^{\mathtt{fake}})$ serves as a penalty term. Furthermore, the toolhit-only terms address partial memorization, providing supplementary evidence to calibrate the final score. Consequently, \name{} ranks these traces using a score $S_\tau$:
    \vspace{-2pt}
\begin{equation}
    S_\tau = Q(R_\tau^{\mathtt{real}}) - \alpha' Q(R_\tau^{\mathtt{fake}}) + \beta' Q(T_\tau^{\mathtt{real}}) - \gamma' Q(T_\tau^{\mathtt{fake}}).
    \label{eq:trace-score}
        \vspace{-2pt}
\end{equation}
where the empirical weighting parameters $\alpha'$, $\beta'$, and $\gamma'$ are detailed in Appendix~\ref{sec:experimental-details}. Finally, the system identifies the top-$K$ traces with the highest $S_\tau$ most likely included in the training set $\mathcal{D}$.
\vspace{-8pt}

\section{Experiments}
    \vspace{-8pt}
\label{sec:experiments}
\subsection{Experimental Setup}
    \vspace{-5pt}
We evaluate \name{} in agent distillation. We supervise thoughts and actions from the trajectories and fine-tune the student models using LoRA. The specific training configurations are rank $32$, learning rate $2\times10^{-4}$, and $\alpha = 64$. Prior traffic-analysis systems have reported identification precision above $70\%$~\cite{chiang2024chatbot, liu2026embarrassingly}. We therefore adopt a more conservative $50\%$ precision setting to avoid missing potential distillation traces, with $|\mathcal{S}|=100$ suspicious trajectories and $|\mathcal{D}_c|=50$ trajectories used for distillation. Appendix~\ref{sec:experimental-details} provides the remaining experimental details, including hyperparameter selection and baselines. Our experiments assess watermark effectiveness, harmlessness, robustness against adversarial processing.

\partitle{Datasets and Models} Our evaluation spans three datasets: BFCL~\cite{patil2025berkeley}, SWE-bench~\cite{jimenez2024swe}, and Telecom (a service-tool domain from the Tau2 benchmark\footnote{\url{https://github.com/sierra-research/tau2-bench}}). GPT-OSS-120B~\cite{agarwal2025gpt} and Kimi-K2.5~\cite{team2026kimi} serve as teacher agents ($\mathcal{M}_t$), with Qwen3-8B acting as both the candidate scorer ($\mathcal{M}_q$) and safety model ($\mathcal{M}_{\mathtt{safe}}$). For distillation, we fine-tune four student models ($\mathcal{M}_s$): Mistral-Small-3.1-24B~\cite{liu2026ministral}, Qwen3-14B~\cite{yang2025qwen3}, Qwen3-32B~\cite{yang2025qwen3}, and GLM-4.7-Flash~\cite{glm2024chatglm}.

\partitle{Evaluation metrics} For detection, we report strict fullhit rates $(R_\tau^{\mathtt{real}},R_\tau^{\mathtt{fake}})$ on real and fake probes, alongside the one-sided $p$-value. A successful detection requires $p<0.05$ and a real-fake mean fullhit gap $\ge 0.05$. For trace-level attribution, we measure $\operatorname{\mathtt{Precision}}@K$ of top-$K$ trajectories belonging to the distillation subset $\mathcal{D}_c$. For harmlessness, we compare task accuracy ($\mathtt{Acc}$) and mean interaction steps ($L$) between control and watermarked agents.
\vspace{-5pt}
\subsection{Effectiveness}
\vspace{-5pt}
\label{sec:detection-effectiveness}

\begin{table*}[htbp]
\centering
\renewcommand{\arraystretch}{0.8}
\setlength{\tabcolsep}{3.5pt}
\caption{True-positive results. Cards is the number of cards; hit-rate cells report hits / probes (rate). $w/\ell$ gives card-level real / fake wins. The `Sig' column indicates statistical significance: {\color{green}$\checkmark$} denotes $p < 0.05$, indicating a statistically significant change, whereas {\color{red}$\times$} denotes $p \ge 0.05$.}
\vspace{-5pt}
\label{tab:true-positive}
\scriptsize
\begin{tabular}{l l c l | c c c c c}
\noalign{\hrule height 1.5pt}
\multirow{2}{*}{\textbf{Teacher Agent$\mathcal{M}_t$}} & \multirow{2}{*}{\textbf{Dataset}} & \multirow{2}{*}{\textbf{Cards}} & \multirow{2}{*}{\textbf{Student Model $\mathcal{M}_s$}} & \multicolumn{5}{c}{\textbf{Detection Results}} \\
\cline{5-9}
& & & & \textbf{$R^{\mathtt{real}}$ (\%)} & \textbf{$R^{\mathtt{fake}}$ (\%)} & \textbf{$w/\ell$} & \textbf{$p$-value} & \textbf{Sig.} \\
\hline
\multirow{4}{*}{GPT-OSS-120B} & \multirow{4}{*}{BFCL} & \multirow{4}{*}{118} & Mistral-24B & \tableRate{93}{182}{51.1} & \tableRate{59}{182}{32.4} & 25 / 1 & \tablePval{4.0}{-7} & {\color{green}\checkmark} \\
& & & GLM-4.7-Flash & \tableRate{108}{182}{59.3} & \tableRate{80}{182}{44.0} & 19 / 1 & \tablePval{2.0}{-5} & {\color{green}\checkmark} \\
& & & Qwen3-14B & \tableRate{44}{182}{24.2} & \tableRate{28}{182}{15.4} & 20 / 4 & \tablePval{7.7}{-4} & {\color{green}\checkmark} \\
& & & Qwen3-32B & \tableRate{89}{182}{48.9} & \tableRate{58}{182}{31.9} & 30 / 9 & \tablePval{5.3}{-4} & {\color{green}\checkmark} \\
\hline
\multirow{4}{*}{Kimi-K2.5} & \multirow{4}{*}{BFCL} & \multirow{4}{*}{145} & Mistral-24B & \tableRate{175}{379}{46.2} & \tableRate{105}{379}{27.7} & 45 / 10 & \tablePval{1.0}{-6} & {\color{green}\checkmark} \\
& & & GLM-4.7-Flash & \tableRate{155}{379}{40.9} & \tableRate{69}{379}{18.2} & 47 / 4 & \tablePval{1.2}{-10} & {\color{green}\checkmark} \\
& & & Qwen3-14B & \tableRate{102}{379}{26.9} & \tableRate{26}{379}{6.9} & 46 / 4 & \tablePval{2.2}{-10} & {\color{green}\checkmark} \\
& & & Qwen3-32B & \tableRate{82}{379}{21.6} & \tableRate{18}{379}{4.7} & 39 / 5 & \tablePval{7.0}{-8} & {\color{green}\checkmark} \\
\noalign{\hrule height 1.5pt}
\end{tabular}
\end{table*}
\vspace{-5pt}

In this section, all student models are trained on the original  dataset $\mathcal{D}_c$. We test our scheme on distilled and clean models via separate true positive and false positive experiments. We also evaluate the performance of existing agent watermark baselines under the matched setting. Furthermore, we evaluate \name{}'s trace-level attribution using $\operatorname{\mathtt{Precision}}@K$ in Appendix~\ref{sec:supplementary-effectiveness}. Finally, we measure how identification accuracy affects statistical significance in Appendix~\ref{sec:supplementary-effectiveness}.
\vspace{-5pt}

\partitle{True positives} We first evaluate whether \name{} can successfully identify student models distilled from watermarked trajectories. Our experiments cover all combinations of 3 datasets (SWE and Telecom are in Appendix~\ref{sec:supplementary-effectiveness}), 2 teacher agents, and 4 student models. In each setting, we compare the hit rates between the real side and the fake side and calculate the $p$-value. The results in the Table~\ref{tab:true-positive} and~\ref{tab:appendix-true-positive} show that \name{} successfully achieves detection in all settings. These results show that \name{} can extract watermark from student models of different architectures.

\partitle{False positives} We evaluate whether \name{} causes false positives on clean models that have not seen watermarked trajectories. As shown in Table~\ref{tab:false-positive} and Table~\ref{tab:appendix-false-positive} in Appendix~\ref{sec:supplementary-effectiveness}, in addition to the four undistilled student models, we also select four base models (DeepSeek-V4-Flash \cite{xu2026deepseek}, MiniMax-M2.5 \cite{chen2026minimax}, Qwen3.5-Flash \cite{yang2025qwen3}, and MiMo-V2.5 \cite{xiao2026mimo}) and build a total of 48 controlled false-positive evaluation settings. The results show that \name{} achieves zero false positives across all 48 settings.

\par\medskip
\noindent
\begin{minipage}{\textwidth}
\refstepcounter{table}
\label{tab:false-positive}
\noindent\textbf{Table~\thetable:} False-positive results. The `Sig' column indicates statistical significance: {\color{green}$\checkmark$} denotes $p < 0.05$, indicating a statistically significant change, whereas {\color{red}$\times$} denotes $p \ge 0.05$.\par\smallskip
\centering
\renewcommand{\arraystretch}{0.8}
\setlength{\tabcolsep}{5.3pt}
\scriptsize
\begin{tabular}{l l c l | c c c c c}
\noalign{\hrule height 1.5pt}
\multirow{2}{*}{\textbf{Teacher Agent$\mathcal{M}_t$}} & \multirow{2}{*}{\textbf{Dataset}} & \multirow{2}{*}{\textbf{Cards}} & \multirow{2}{*}{\textbf{Clean Model $\mathcal{M}_c$}} & \multicolumn{5}{c}{\textbf{Detection Results}} \\
\cline{5-9}
& & & & \textbf{$R^{\mathtt{real}}$ (\%)} & \textbf{$R^{\mathtt{fake}}$ (\%)} & \textbf{$w/\ell$} & \textbf{$p$-value} & \textbf{Sig.} \\
\hline
\multirow{8}{*}{GPT-OSS-120B} & \multirow{8}{*}{BFCL} & \multirow{8}{*}{118} & DeepSeek-V4-Flash & \tableRate{10}{182}{5.5} & \tableRate{11}{182}{6.0} & 2 / 3 & \tablePval{8.1}{-1} & {\color{red}$\times$} \\
& & & MiniMax-M2.5 & \tableRate{9}{182}{4.9} & \tableRate{10}{182}{5.5} & 2 / 3 & \tablePval{8.1}{-1} & {\color{red}$\times$} \\
& & & Mistral-24B & \tableRate{9}{182}{4.9} & \tableRate{7}{182}{3.8} & 3 / 2 & \tablePval{5.0}{-1} & {\color{red}$\times$} \\
& & & Qwen3-14B & \tableRate{7}{182}{3.8} & \tableRate{6}{182}{3.3} & 2 / 1 & \tablePval{5.0}{-1} & {\color{red}$\times$} \\
& & & Qwen3-32B & \tableRate{4}{182}{2.2} & \tableRate{4}{182}{2.2} & 2 / 2 & \tablePval{6.9}{-1} & {\color{red}$\times$} \\
& & & Qwen3.5-Flash & \tableRate{8}{182}{4.4} & \tableRate{8}{182}{4.4} & 2 / 1 & \tablePval{5.0}{-1} & {\color{red}$\times$} \\
& & & MiMo-V2.5 & \tableRate{10}{182}{5.5} & \tableRate{11}{182}{6.0} & 2 / 2 & \tablePval{6.9}{-1} & {\color{red}$\times$} \\
& & & GLM-4.7-Flash & \tableRate{5}{182}{2.7} & \tableRate{1}{182}{0.5} & 4 / 0 & \tablePval{6.2}{-2} & {\color{red}$\times$} \\
\hline
\multirow{8}{*}{Kimi-K2.5} & \multirow{8}{*}{BFCL} & \multirow{8}{*}{145} & DeepSeek-V4-Flash & \tableRate{10}{379}{2.6} & \tableRate{9}{379}{2.4} & 5 / 4 & \tablePval{5.0}{-1} & {\color{red}$\times$} \\
& & & MiniMax-M2.5 & \tableRate{6}{379}{1.6} & \tableRate{3}{379}{0.8} & 4 / 2 & \tablePval{3.4}{-1} & {\color{red}$\times$} \\
& & & Mistral-24B & \tableRate{1}{379}{0.3} & \tableRate{3}{379}{0.8} & 1 / 2 & \tablePval{8.8}{-1} & {\color{red}$\times$} \\
& & & Qwen3-14B & \tableRate{1}{379}{0.3} & \tableRate{0}{379}{0.0} & 1 / 0 & \tablePval{5.0}{-1} & {\color{red}$\times$} \\
& & & Qwen3-32B & \tableRate{1}{379}{0.3} & \tableRate{0}{379}{0.0} & 1 / 0 & \tablePval{5.0}{-1} & {\color{red}$\times$} \\
& & & Qwen3.5-Flash & \tableRate{0}{379}{0.0} & \tableRate{0}{379}{0.0} & 0 / 0 & \tablePval{1.0}{0} & {\color{red}$\times$} \\
& & & MiMo-V2.5 & \tableRate{6}{379}{1.6} & \tableRate{6}{379}{1.6} & 4 / 3 & \tablePval{5.0}{-1} & {\color{red}$\times$} \\
& & & GLM-4.7-Flash & \tableRate{2}{379}{0.5} & \tableRate{2}{379}{0.5} & 2 / 1 & \tablePval{5.0}{-1} & {\color{red}$\times$} \\
\noalign{\hrule height 1.5pt}
\end{tabular}
\end{minipage}

\vspace{-5pt}
\par\medskip
\begin{wraptable}[10]{l}{0.52\textwidth}
\vspace{-\intextsep}
\refstepcounter{table}
\label{tab:baseline-effectiveness}
\noindent\textbf{Table~\thetable:} Baseline effectiveness. Agentwm is significant at $\geq3/5$ passes; SeqWM at median $p<0.05$.\par\smallskip
\scriptsize
\renewcommand{\arraystretch}{0.9}
\setlength{\tabcolsep}{3pt}
\begin{tabular}{l l | c c | c c}
\noalign{\hrule height 1.5pt}
\multirow{2}{*}{\textbf{Dataset}} & \multirow{2}{*}{\textbf{Student Model $\mathcal{M}_s$}} & \multicolumn{2}{c|}{\textbf{Agentwm}} & \multicolumn{2}{c}{\textbf{SeqWM}} \\
\cline{3-6}
& & \textbf{Passes} & \textbf{Sig.} & \textbf{Median $p$} & \textbf{Sig.} \\
\hline
\multirow{4}{*}{BFCL} & Mistral-24B & 1/5 & {\color{red}$\times$} & 0.6848 & {\color{red}$\times$} \\
& GLM-4.7-Flash & 3/5 & {\color{green}\checkmark} & 0.5130 & {\color{red}$\times$} \\
& Qwen3-14B & 1/5 & {\color{red}$\times$} & 0.4650 & {\color{red}$\times$} \\
& Qwen3-32B & 2/5 & {\color{red}$\times$} & 0.5744 & {\color{red}$\times$} \\
\cline{1-6}
\multirow{4}{*}{SWE-bench} & Mistral-24B & 1/5 & {\color{red}$\times$} & 0.6274 & {\color{red}$\times$} \\
& GLM-4.7-Flash & 3/5 & {\color{green}\checkmark} & 0.5914 & {\color{red}$\times$} \\
& Qwen3-14B & 1/5 & {\color{red}$\times$} & 0.7073 & {\color{red}$\times$} \\
& Qwen3-32B & 0/5 & {\color{red}$\times$} & 0.4895 & {\color{red}$\times$} \\
\noalign{\hrule height 1.5pt}
\end{tabular}
\vspace{-\intextsep}
\end{wraptable}

\partitle{Baseline effectiveness} We evaluate the effectiveness of the two baseline schemes under the matched distillation setting and the details about baselines are in Appendix~\ref{sec:experimental-details}. As shown in Table~\ref{tab:baseline-effectiveness}, for Agentwm, only the GLM-4.7-Flash student model meets the passing criterion of $\geq 3/5$ passes, while all other students fail. For Seqwm, the median $p$-values across all tested models are greater than 0.05, failing to successfully detect any distilled models. As expected, baselines fail, because Seqwm is not designed for antidistillation, and Agentwm has weak watermark signals that requires identical teacher and student base models.
\vspace{-5pt}
\subsection{Harmlessness}
\label{sec:harmlessness}
\begin{figure}[H]
\centering
\begin{subfigure}[t]{0.242\textwidth}
\centering
\includegraphics[width=\linewidth]{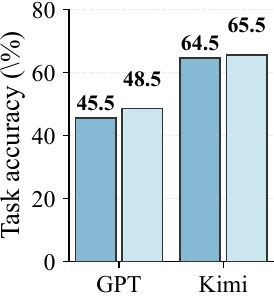}
\caption{BFCL: accuracy}
\end{subfigure}\hfill
\begin{subfigure}[t]{0.242\textwidth}
\centering
\includegraphics[width=\linewidth]{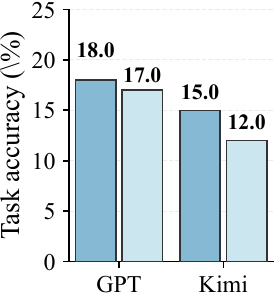}
\caption{Telecom: accuracy}
\end{subfigure}\hfill
\begin{subfigure}[t]{0.242\textwidth}
\centering
\includegraphics[width=\linewidth]{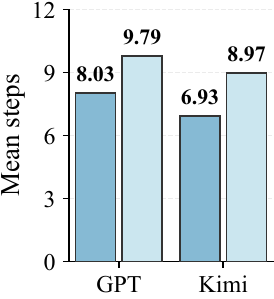}
\caption{BFCL: steps}
\end{subfigure}\hfill
\begin{subfigure}[t]{0.242\textwidth}
\centering
\includegraphics[width=\linewidth]{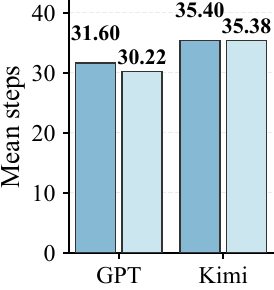}
\caption{Telecom: steps}
\end{subfigure}
\caption{
Harmlessness results. In (a) and (b), the vertical axis reports accuracy; in (c) and (d), it reports the mean number of steps. Colors distinguish \textcolor[HTML]{86BAD4}{$\blacksquare$} unwatermarked and \textcolor[HTML]{CBE6EF}{$\blacksquare$} watermarked runs.
}
\vspace{-15pt}
\label{fig:harmlessness}
\end{figure}

\vspace{-5pt}
\partitle{Harmlessness} A reasonable watermark scheme should not noticeably affect the performance of the teacher agent. Therefore, we evaluate impact of \name{} on the teacher agent $\mathcal{M}_t$'s task utility and inference cost. For the four cohorts in BFCL and Telecom, we compare the unwatermarked agent $\mathcal{M}_\mathtt{tc}$ and the watermarked teacher agent $\mathcal{M}_t$ on 200 tasks. We measure the task accuracy difference $\Delta \mathtt{Acc} = \mathbb{E}[\mathtt{Acc}(\mathcal{M}_t) - \mathtt{Acc}(\mathcal{M}_\mathtt{tc})]$ and the expected interaction step difference $\Delta L = \mathbb{E}[L(\mathcal{M}_t) - L(\mathcal{M}_\mathtt{tc})]$. As summarized in Figure~\ref{fig:harmlessness}, embedding the watermark does not cause a drop in overall performance. The task accuracy difference $\Delta \mathtt{Acc}$ ranges only from $-3.0\%$ to $+3.0\%$, indicating that the accuracy loss is acceptable. Meanwhile, the increase in interaction cost is acceptable: the average step count increases from 20.49 to 21.09 steps, and $\Delta L$ ranges from $-1.38$ to $+2.04$ steps.
\vspace{-3pt}
\FloatBarrier

\raggedbottom
\vspace{-5pt}
\subsection{Robustness}
\label{sec:robustness}
\vspace{-5pt}
To demonstrate the robustness of our watermark scheme, We evaluate \name{}'s robustness against four common pre-distillation data transformations. These operations include: data flooding, paraphrasing attack, adaptive attack, advanced data flooding and truncation (the last two are in Appendix~\ref{sec:supplementary-robustness}). We evaluate the robustness of Agentwm under data flooding in Appendix~\ref{sec:supplementary-robustness}. We use GPT-OSS-120B as the teacher agent.

\begin{figure}[H]
\centering
\begin{subfigure}[t]{\robustnessPanelWidth}
\centering
\includegraphics[width=0.9\linewidth]{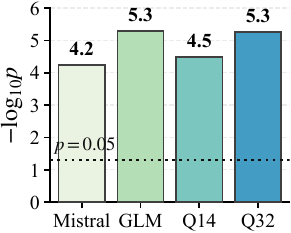}
\caption{BFCL: D1}
\end{subfigure}\hfill
\begin{subfigure}[t]{\robustnessPanelWidth}
\centering
\includegraphics[width=0.9\linewidth]{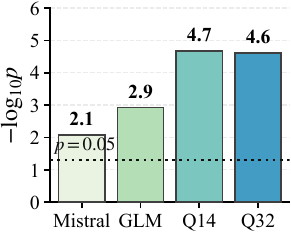}
\caption{BFCL: D5}
\end{subfigure}\hfill
\begin{subfigure}[t]{\robustnessPanelWidth}
\centering
\includegraphics[width=0.9\linewidth]{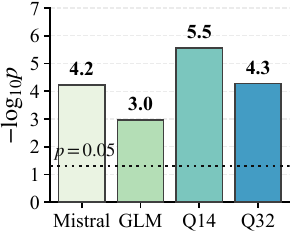}
\caption{BFCL: D10}
\end{subfigure}\par\vspace{-0.35em}
\begin{subfigure}[t]{\robustnessPanelWidth}
\centering
\includegraphics[width=0.9\linewidth]{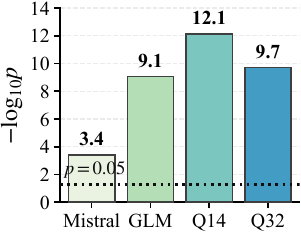}
\caption{SWE-bench: D1}
\end{subfigure}\hfill
\begin{subfigure}[t]{\robustnessPanelWidth}
\centering
\includegraphics[width=0.9\linewidth]{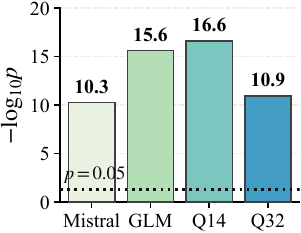}
\caption{SWE-bench: D5}
\end{subfigure}\hfill
\begin{subfigure}[t]{\robustnessPanelWidth}
\centering
\includegraphics[width=0.9\linewidth]{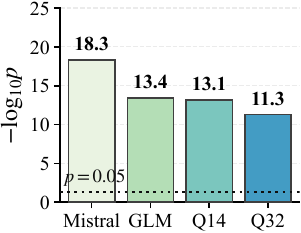}
\caption{SWE-bench: D10}
\end{subfigure}
\vspace{-5pt}
\caption{Data-flooding robustness. The vertical axis reports $-\log_{10}p$. In (a)--(f), \robustnessModelLegend. The \textcolor[HTML]{8C8C8C}{\hdashrule[0.5ex]{1.2em}{1.2pt}{1pt 1.5pt}} marks $p=0.05$.}
\vspace{-25pt}
\label{fig:data-flooding-robustness}
\end{figure}

\partitle{Data flooding} Before distillation, an attacker may mix the training set with standard clean data to improve distillation performance~\cite{ouyang2022training, mukherjee2023orca}, which will dilutes the watermark signals. To evaluate its impact on our watermark strength, we mix the original watermarked dataset $\mathcal{D}_c$ with clean standard answers. Specifically, in the D1, D5, and D10 settings, we incorporate $1\times$, $5\times$, and $10\times$ the amount of clean data relative to $\mathcal{D}_c$. We evaluate a total of 24 settings across 2 benchmarks (BFCL and SWE-bench), 3 dilution ratios, and 4 student models. As shown in Figure~\ref{fig:data-flooding-robustness}, \name{} successfully detects every setting. Even under the strongest $1{:}10$ dilution ratio, all 8 tested student models remain successfully detected.

\par\medskip
\begin{wraptable}{l}{0.45\textwidth}
\vspace{-\intextsep}
\refstepcounter{table}
\label{tab:semantic-rewriting-robustness}
\noindent\textbf{Table~\thetable:} Paraphrasing attack robustness. 
\centering
\scriptsize
\setlength{\tabcolsep}{2pt}
\begin{tabular*}{\linewidth}{@{\extracolsep{\fill}}l l | c c@{}}
\noalign{\hrule height 1.5pt}
\textbf{Dataset} & \textbf{Student Model $\mathcal{M}_s$} & \textbf{$p$-value} & \textbf{Sig.} \\
\hline
\multirow{4}{*}{BFCL} & Mistral-24B & \tablePval{5.7}{-5} & {\color{green}\checkmark} \\
& GLM-4.7-Flash & \tablePval{1.4}{-4} & {\color{green}\checkmark} \\
& Qwen3-14B & \tablePval{3.3}{-3} & {\color{green}\checkmark} \\
& Qwen3-32B & \tablePval{9.8}{-5} & {\color{green}\checkmark} \\
\hline
\multirow{4}{*}{Telecom} & Mistral-24B & \tablePval{3.2}{-2} & {\color{green}\checkmark} \\
& GLM-4.7-Flash & \tablePval{1.4}{-5} & {\color{green}\checkmark} \\
& Qwen3-14B & \tablePval{3.5}{-3} & {\color{green}\checkmark} \\
& Qwen3-32B & \tablePval{4.1}{-3} & {\color{green}\checkmark} \\
\noalign{\hrule height 1.5pt}
\end{tabular*}
\vspace{-\intextsep}
\end{wraptable}
\partitle{Paraphrasing attack} An attacker may paraphrase the trajectories, aiming to alter semantic patterns and evade watermark detection while preserving the utility~\cite{krishna2023paraphrasing}. To evaluate the robustness against this rewriting attack, we use DeepSeek-V4-Flash to paraphrase the  $\mathcal{D}_c$. Subsequently, we train 4 student models on these rewritten datasets across the BFCL and Telecom and detect their watermark signals. As reported in Table~\ref{tab:semantic-rewriting-robustness}, \name{} successfully detects the watermark across all 8 settings showing our scheme's robustness under paraphrasing.

\par\medskip
\begin{wraptable}{l}{0.45\textwidth}
\vspace{-\intextsep}
\refstepcounter{table}
\label{tab:adaptive-attack}
\noindent\textbf{Table~\thetable:} Adaptive attack robustness.\par\smallskip
\centering
\scriptsize
\setlength{\tabcolsep}{2pt}
\begin{tabular*}{\linewidth}{@{\extracolsep{\fill}}l l | c c@{}}
\noalign{\hrule height 1.5pt}
\textbf{Dataset} & \textbf{Student Model $\mathcal{M}_s$} & \textbf{$p$-value} & \textbf{Sig.} \\
\hline
\multirow{4}{*}{BFCL} & Mistral-24B & \tablePval{9.2}{-4} & {\color{green}\checkmark} \\
& GLM-4.7-Flash & \tablePval{1.4}{-4} & {\color{green}\checkmark} \\
& Qwen3-14B & \tablePval{9.6}{-3} & {\color{green}\checkmark} \\
& Qwen3-32B & \tablePval{4.1}{-4} & {\color{green}\checkmark} \\
\hline
\multirow{4}{*}{Telecom} & Mistral-24B & \tablePval{2.8}{-8} & {\color{green}\checkmark} \\
& GLM-4.7-Flash & \tablePval{3.7}{-3} & {\color{green}\checkmark} \\
& Qwen3-14B & \tablePval{1.0}{-11} & {\color{green}\checkmark} \\
& Qwen3-32B & \tablePval{5.5}{-12} & {\color{green}\checkmark} \\
\noalign{\hrule height 1.5pt}
\end{tabular*}
\vspace{-\intextsep}
\end{wraptable}
\partitle{Adaptive attack} In this section, we consider a stronger attacker who knows that our watermark relies on auxiliary behaviors. To avoid detection, the attacker uses DeepSeek-V4-Flash to clean the collected data by finding and removing suspected auxiliary action while keeping the trajectories useful. Next, we fine-tune student models on this cleaned data and test them on BFCL and Telecom. As shown in Table~\ref{tab:adaptive-attack}, even with this removal, \name{} successfully detects all student models (with the minimal $p$-value of $9.6\times10^{-3}$). This shows that our watermark is highly robust, even when an attacker actively tries to move the signals.

\FloatBarrier
\flushbottom
\vspace{-8pt}

\section{Conclusion}
\label{sec:conclusion}
\vspace{-8pt}
We present \name{}, a behavioral watermarking framework to track the unauthorized distillation of LLM agents. \name{} inserts useful extra tool calls after normal core actions. To verify if the watermark is copied, it uses paired real and fake probes along with a card-level sign test. This design supports both model-level detection and trace-level tracking.In experiments across three benchmarks, two teacher models, and four student architectures, \name{} successfully detects all 24 distilled students with zero false positives on 48 clean models. It also remains highly effective under data flooding, rewriting, truncation, and adaptive cleaning attacks. Future work will extend this evaluation to more teacher models, cleaning tools, and training methods.

\section*{AI use statement}
In this work, we used generative AI tools for research execution, generating synthetic datasets, and code generation. Specifically, large language models were employed to generate agent-interaction trajectories, construct paired verification probes for our evaluations, and assist with implementation. We have not used generative AI tools for drafting core sections of the paper or proving mathematical claims, and the rest of the required disclosure tasks are not applicable to this work. Additionally, we used generative AI tools (Gemini 3.1 Pro) for aiding and polishing the English writing of the manuscript.
We have reviewed all AI-assisted work. The authors manually assessed AI-assisted research ideas through a literature survey and technical discussion. For the synthetic datasets, the LLM-generated agent trajectories and verification probes were programmatically validated and manually sampled to ensure they form valid, executable tool-use sequences that strictly adhere to the predefined tool schemas and task objectives. All AI-generated code was reviewed, tested, and, where necessary, revised by the authors before use. For the writing assistance, all AI-polished text was thoroughly reviewed and edited by the authors to guarantee it accurately reflects our original scientific intent, methodology, and conclusions without introducing hallucinations or overclaims. We take responsibility for the final content of this work, including text, claims, or artifacts produced with the aid of generative AI.

\section*{Reproducibility Statement}

To ensure the full reproducibility of our results, we have made our complete source code, benchmark adapters, and evaluation scripts publicly available via an anonymous repository (\href{https://github.com/qx041609/Auxmark}{https://github.com/qx041609/Auxmark}). Additionally, all trained student model weights (e.g., LoRA adapters) necessary to replicate our detection and trace-level attribution experiments are hosted on Hugging Face (\href{https://huggingface.co/AuxMark/AuxMark/tree/main}{huggingface.co/AuxMark/AuxMark/tree/main}).

\bibliography{09_reference}
\bibliographystyle{iclr2027_conference}

\appendix

\appendix


\section{Related Work}

\subsection{Model Distillation}
Knowledge distillation is originally proposed as to transfer the capicity of a teacher model to a student model \cite{hinton2015distilling}. With rapid development of LLM, fine-tuning on teacher-generated data has become an increasingly important form of distillation, allowing student models to acquire advanced capabilities directly from the teacher's responses \cite{hsieh2023distilling, mukherjee2023orca}. This paradigm has also been extended to agent systems \cite{kang2026distilling, liu2026structured, luo2026agentark}. Agent Distillation \cite{kang2026distilling} proposed an agent distillation method that enables student model to learn both tool-use behaviors and reasoning capabilities. AgentArk \cite{luo2026agentark} introduced a method that distills multi-agent dynamics into the weights of a single model. Agent trajectories record not only generated text but also tool use and multi-step decisions, making them valuable supervision for distillation. As these trajectories can be used to cheaply transfer the capabilities of teacher agents, unauthorized distillation poses a growing threat to model copyright and calls for effective detection and tracing mechanisms.

\subsection{Model Watermarking}

\partitle{Content watermarking}
Recent studies have proposed a series of content watermarking methods to identify whether a given output is generated by a specific model. Early methods mainly embed statistical signals during token generation, such as the red-green list mechanism \cite{kirchenbauer2023watermark}. Later studies further improve the robustness and generation quality of watermarks by using semantic information \cite{liu2024semantic, huo2026pmark, huo2026samark,hou2024semstamp}. However, with the development of LLM agents, traditional content watermarks are difficult to directly apply to agent settings. These methods usually embed watermark signals by changing token choices or semantic expressions in free-form text, while the key outputs of agents are often tool calls with strict format requirements. Directly modifying these outputs may break the tool calls and affect correct execution. Therefore, recent studies have started to move watermarks from generated content to agent trajectories and behavior patterns \cite{huang2025agent, an2026sequential, zhang2026memmark}.

\partitle{Model watermarking against distillation}
Several watermarking methods have been proposed to protect models against unauthorized distillation. Methods such as Lexical Watermark and CATER insert special word patterns into teacher outputs, so that student models trained on these outputs can inherit detectable features~\cite{he2022protecting, he2022cater, yang2026asking}. Recent work further targets LLM distillation by changing token generation or rewriting reasoning traces, either reducing the value of teacher outputs for distillation or leaving detectable signals in the student model~\cite{savani2026antidistillation, xu2026antidistillation, ma2026protecting}. However, these methods mainly work on text tokens or reasoning traces, and are difficult to directly apply to agents whose behavior is mainly represented by structured tool calls and multi-step interactions. Recent work has extended anti-distillation watermarking to agent trajectories~\cite{wang2026protecting}, but its main evaluation still focuses on settings where the teacher and student models use the same base model, and its robustness to data processing such as data mixing remains limited. Therefore, a more general and robust watermarking method is still needed for tracing agent distillation.

\section{Supplement of Experiments}

\subsection{Experimental Details}
\label{sec:experimental-details}

\partitle{Hyperparameter selection} We select \name{} hyperparameters through preliminary experiments to balance stealthiness, utility and detection reliability. For dynamic scheduling, we set $p_0 = 0.1$, $\Delta p = 0.1$, $p_{\mathtt{max}} = 0.5$, and the maximum budget $B = 10$. The system generates $K = 3$ candidates, and the scoring weights in Eq.~\ref{equ:score} are $\alpha = 0.35$, $\beta = 0.25$, and $\gamma = 0.40$. The repeatability score is $s_{\mathtt{rep}}^j=0$ for $n_j=0$, $0.6$ for $1\leq n_j<5$, and $1$ for $n_j\geq5$. For detection, we generate $N = 3$ parameterized variants per card and use a significance level $p=0.05$ with margin threshold $r=0.05$. For trace-level attribution in Eq.~\ref{eq:trace-score}, we set $\alpha' = 0.25$, $\beta' = 0.10$, and $\gamma' = 0.10$.

\enlargethispage{\baselineskip}
\partitle{Baselines} We evaluate our approach against two representative agent watermarking baselines. We exclude large language model antidistillation watermarks~\cite{xu2026antidistillation,savani2026antidistillation} from our comparison, as agent environments demand formatted outputs and modifying tokens or reasoning traces can cause format corruption and execution failures. Thus, we focus on agent schemes~\cite{an2026sequential,huang2025agent,luo2026agentark}. We select Seqwm~\cite{an2026sequential} and Agentwm~\cite{wang2026protecting} because Seqwm demonstrates better robustness among trace watermarks generated for agents, and Agentwm specifically targets agent anti-distillation. Their available implementations use different watermark encodings, test units, and decision statistics from \name{}; we therefore report each baseline in its native metric. Specifically, we train the student models $\mathcal{M}_s$ using the distillation dataset $\mathcal{D}_c$ watermarked by each respective scheme, and then have them replay 100 tasks from the set $\mathcal{S}$. For Agentwm, we collect unwatermarked traces using four distinct models to estimate the natural initial distribution $P_c$ of synonymous tool sets. The scheme then modifies this $P_c$ into a watermarked distribution $P_{\mathtt{wm}}$. During detection, it extracts 5 fixed synonymous tool sets (referred to as passes) from the 100 task replays. A pass is considered successfully detected only when the JSD between the student's empirical distribution and the distribution $P_{\mathtt{wm}}$ is at most 0.015. The student model is flagged as significant only if at least 3 out of 5 passes are detected ($\geq 3/5$). For Seqwm, it evaluates each replayed trace against a background null distribution generated from 1,000 incorrect keys. It uses the median $p$-value of these 100 test traces to determine whether the student model $\mathcal{M}_s$ is distilled, requiring a median $p < 0.05$ for a successful detection.

\subsection{Supplementary Effectiveness Results}
\label{sec:supplementary-effectiveness}
\begin{table*}[ht]
\centering
\renewcommand{\arraystretch}{0.8}
\setlength{\tabcolsep}{3.5pt}
\caption{Supplementary true-positive results for SWE-bench and Telecom.
Cards is the number of cards; hit-rate cells report hits / probes (rate).
$w/\ell$ gives card-level real / fake wins.
The Sig. column indicates statistical significance:
{\color{green}$\checkmark$} denotes $p < 0.05$,
whereas {\color{red}$\times$} denotes $p \ge 0.05$.}
\label{tab:appendix-true-positive}

\scriptsize
\begin{tabular}{l l c l | c c c c c}
\noalign{\hrule height 1.5pt}
\multirow{2}{*}{\textbf{Teacher Agent$\mathcal{M}_t$}} & \multirow{2}{*}{\textbf{Dataset}} & \multirow{2}{*}{\textbf{Cards}} & \multirow{2}{*}{\textbf{Student Model $\mathcal{M}_s$}} & \multicolumn{5}{c}{\textbf{Detection Results}} \\
\cline{5-9}
& & & & \textbf{$R^{\mathtt{real}}$ (\%)} & \textbf{$R^{\mathtt{fake}}$ (\%)} & \textbf{$w/\ell$} & \textbf{$p$-value} & \textbf{Sig.} \\
\hline
\multirow{8}{*}{GPT-OSS-120B} & \multirow{4}{*}{SWE-bench} & \multirow{4}{*}{274} & Mistral-24B & \tableRate{60}{387}{15.5} & \tableRate{18}{387}{4.7} & 46 / 6 & \tablePval{5.2}{-9} & {\color{green}\checkmark} \\
& & & GLM-4.7-Flash & \tableRate{81}{387}{20.9} & \tableRate{21}{387}{5.4} & 67 / 6 & \tablePval{2.0}{-14} & {\color{green}\checkmark} \\
& & & Qwen3-14B & \tableRate{67}{387}{17.3} & \tableRate{18}{387}{4.7} & 54 / 4 & \tablePval{1.6}{-12} & {\color{green}\checkmark} \\
& & & Qwen3-32B & \tableRate{72}{387}{18.6} & \tableRate{20}{387}{5.2} & 58 / 6 & \tablePval{4.5}{-12} & {\color{green}\checkmark} \\
\cline{2-9}
& \multirow{4}{*}{Telecom} & \multirow{4}{*}{125} & Mistral-24B & \tableRate{83}{146}{56.8} & \tableRate{46}{146}{31.5} & 44 / 7 & \tablePval{6.1}{-8} & {\color{green}\checkmark} \\
& & & GLM-4.7-Flash & \tableRate{76}{146}{52.1} & \tableRate{42}{146}{28.8} & 42 / 4 & \tablePval{2.6}{-9} & {\color{green}\checkmark} \\
& & & Qwen3-14B & \tableRate{86}{146}{58.9} & \tableRate{44}{146}{30.1} & 48 / 6 & \tablePval{1.6}{-9} & {\color{green}\checkmark} \\
& & & Qwen3-32B & \tableRate{85}{146}{58.2} & \tableRate{36}{146}{24.7} & 57 / 7 & \tablePval{3.8}{-11} & {\color{green}\checkmark} \\
\hline
\multirow{8}{*}{Kimi-K2.5} & \multirow{4}{*}{SWE-bench} & \multirow{4}{*}{262} & Mistral-24B & \tableRate{85}{437}{19.5} & \tableRate{50}{437}{11.4} & 38 / 9 & \tablePval{1.2}{-5} & {\color{green}\checkmark} \\
& & & GLM-4.7-Flash & \tableRate{80}{437}{18.3} & \tableRate{42}{437}{9.6} & 34 / 5 & \tablePval{1.2}{-6} & {\color{green}\checkmark} \\
& & & Qwen3-14B & \tableRate{48}{437}{11.0} & \tableRate{17}{437}{3.9} & 33 / 8 & \tablePval{5.6}{-5} & {\color{green}\checkmark} \\
& & & Qwen3-32B & \tableRate{45}{437}{10.3} & \tableRate{14}{437}{3.2} & 27 / 1 & \tablePval{1.1}{-7} & {\color{green}\checkmark} \\
\cline{2-9}
& \multirow{4}{*}{Telecom} & \multirow{4}{*}{122} & Mistral-24B & \tableRate{68}{192}{35.4} & \tableRate{48}{192}{25.0} & 37 / 14 & \tablePval{8.8}{-4} & {\color{green}\checkmark} \\
& & & GLM-4.7-Flash & \tableRate{56}{192}{29.2} & \tableRate{28}{192}{14.6} & 33 / 6 & \tablePval{7.1}{-6} & {\color{green}\checkmark} \\
& & & Qwen3-14B & \tableRate{60}{192}{31.2} & \tableRate{36}{192}{18.8} & 33 / 9 & \tablePval{1.4}{-4} & {\color{green}\checkmark} \\
& & & Qwen3-32B & \tableRate{66}{192}{34.4} & \tableRate{43}{192}{22.4} & 27 / 6 & \tablePval{1.6}{-4} & {\color{green}\checkmark} \\
\noalign{\hrule height 1.5pt}
\end{tabular}
\end{table*}

\begin{table*}[ht]
\centering
\renewcommand{\arraystretch}{0.8}
\setlength{\tabcolsep}{3.5pt}
\caption{Supplementary false-positive results for SWE-bench and Telecom.
Cards is the number of cards; hit-rate cells report hits / probes (rate).
$w/\ell$ gives card-level real / fake wins.
The Sig. column indicates statistical significance:
{\color{green}$\checkmark$} denotes $p < 0.05$,
whereas {\color{red}$\times$} denotes $p \ge 0.05$.}
\label{tab:appendix-false-positive}

\scriptsize
\begin{tabular}{l l c l | c c c c c}
\noalign{\hrule height 1.5pt}
\multirow{2}{*}{\textbf{Teacher Agent$\mathcal{M}_t$}} & \multirow{2}{*}{\textbf{Dataset}} & \multirow{2}{*}{\textbf{Cards}} & \multirow{2}{*}{\textbf{Clean Model $\mathcal{M}_c$}} & \multicolumn{5}{c}{\textbf{Detection Results}} \\
\cline{5-9}
& & & & \textbf{$R^{\mathtt{real}}$ (\%)} & \textbf{$R^{\mathtt{fake}}$ (\%)} & \textbf{$w/\ell$} & \textbf{$p$-value} & \textbf{Sig.} \\
\hline
\multirow{16}{*}{GPT-OSS-120B} & \multirow{8}{*}{SWE-bench} & \multirow{8}{*}{274} & DeepSeek-V4-Flash & \tableRate{0}{387}{0.0} & \tableRate{0}{387}{0.0} & 0 / 0 & \tablePval{1.0}{0} & {\color{red}$\times$} \\
& & & MiniMax-M2.5 & \tableRate{5}{387}{1.3} & \tableRate{2}{387}{0.5} & 4 / 1 & \tablePval{1.9}{-1} & {\color{red}$\times$} \\
& & & Mistral-24B & \tableRate{15}{387}{3.9} & \tableRate{10}{387}{2.6} & 10 / 6 & \tablePval{2.3}{-1} & {\color{red}$\times$} \\
& & & Qwen3-14B & \tableRate{0}{387}{0.0} & \tableRate{3}{387}{0.8} & 0 / 3 & \tablePval{1.0}{0} & {\color{red}$\times$} \\
& & & Qwen3-32B & \tableRate{0}{387}{0.0} & \tableRate{2}{387}{0.5} & 0 / 2 & \tablePval{1.0}{0} & {\color{red}$\times$} \\
& & & Qwen3.5-Flash & \tableRate{4}{387}{1.0} & \tableRate{5}{387}{1.3} & 4 / 3 & \tablePval{5.0}{-1} & {\color{red}$\times$} \\
& & & MiMo-V2.5 & \tableRate{3}{387}{0.8} & \tableRate{3}{387}{0.8} & 2 / 2 & \tablePval{6.9}{-1} & {\color{red}$\times$} \\
& & & GLM-4.7-Flash & \tableRate{1}{387}{0.3} & \tableRate{2}{387}{0.5} & 1 / 2 & \tablePval{8.8}{-1} & {\color{red}$\times$} \\
\cline{2-9}
& \multirow{8}{*}{Telecom} & \multirow{8}{*}{125} & DeepSeek-V4-Flash & \tableRate{18}{146}{12.3} & \tableRate{13}{146}{8.9} & 10 / 5 & \tablePval{1.5}{-1} & {\color{red}$\times$} \\
& & & MiniMax-M2.5 & \tableRate{21}{146}{14.4} & \tableRate{19}{146}{13.0} & 6 / 3 & \tablePval{2.5}{-1} & {\color{red}$\times$} \\
& & & Mistral-24B & \tableRate{11}{146}{7.5} & \tableRate{9}{146}{6.2} & 6 / 5 & \tablePval{5.0}{-1} & {\color{red}$\times$} \\
& & & Qwen3-14B & \tableRate{17}{146}{11.6} & \tableRate{21}{146}{14.4} & 3 / 6 & \tablePval{9.1}{-1} & {\color{red}$\times$} \\
& & & Qwen3-32B & \tableRate{17}{146}{11.6} & \tableRate{18}{146}{12.3} & 2 / 3 & \tablePval{8.1}{-1} & {\color{red}$\times$} \\
& & & Qwen3.5-Flash & \tableRate{11}{146}{7.5} & \tableRate{8}{146}{5.5} & 6 / 3 & \tablePval{2.5}{-1} & {\color{red}$\times$} \\
& & & MiMo-V2.5 & \tableRate{13}{146}{8.9} & \tableRate{13}{146}{8.9} & 8 / 8 & \tablePval{6.0}{-1} & {\color{red}$\times$} \\
& & & GLM-4.7-Flash & \tableRate{1}{146}{0.7} & \tableRate{0}{146}{0.0} & 1 / 0 & \tablePval{5.0}{-1} & {\color{red}$\times$} \\
\hline
\multirow{16}{*}{Kimi-K2.5} & \multirow{8}{*}{SWE-bench} & \multirow{8}{*}{262} & DeepSeek-V4-Flash & \tableRate{12}{437}{2.7} & \tableRate{5}{437}{1.1} & 8 / 3 & \tablePval{1.1}{-1} & {\color{red}$\times$} \\
& & & MiniMax-M2.5 & \tableRate{14}{437}{3.2} & \tableRate{7}{437}{1.6} & 9 / 4 & \tablePval{1.3}{-1} & {\color{red}$\times$} \\
& & & Mistral-24B & \tableRate{20}{437}{4.6} & \tableRate{11}{437}{2.5} & 10 / 4 & \tablePval{9.0}{-2} & {\color{red}$\times$} \\
& & & Qwen3-14B & \tableRate{1}{437}{0.2} & \tableRate{1}{437}{0.2} & 1 / 1 & \tablePval{7.5}{-1} & {\color{red}$\times$} \\
& & & Qwen3-32B & \tableRate{2}{437}{0.5} & \tableRate{2}{437}{0.5} & 2 / 2 & \tablePval{6.9}{-1} & {\color{red}$\times$} \\
& & & Qwen3.5-Flash & \tableRate{15}{437}{3.4} & \tableRate{13}{437}{3.0} & 6 / 4 & \tablePval{3.8}{-1} & {\color{red}$\times$} \\
& & & MiMo-V2.5 & \tableRate{10}{437}{2.3} & \tableRate{6}{437}{1.4} & 8 / 5 & \tablePval{2.9}{-1} & {\color{red}$\times$} \\
& & & GLM-4.7-Flash & \tableRate{5}{437}{1.1} & \tableRate{1}{437}{0.2} & 4 / 0 & \tablePval{6.2}{-2} & {\color{red}$\times$} \\
\cline{2-9}
& \multirow{8}{*}{Telecom} & \multirow{8}{*}{122} & DeepSeek-V4-Flash & \tableRate{11}{192}{5.7} & \tableRate{8}{192}{4.2} & 4 / 1 & \tablePval{1.9}{-1} & {\color{red}$\times$} \\
& & & MiniMax-M2.5 & \tableRate{23}{192}{12.0} & \tableRate{18}{192}{9.4} & 9 / 5 & \tablePval{2.1}{-1} & {\color{red}$\times$} \\
& & & Mistral-24B & \tableRate{19}{192}{9.9} & \tableRate{16}{192}{8.3} & 14 / 11 & \tablePval{3.5}{-1} & {\color{red}$\times$} \\
& & & Qwen3-14B & \tableRate{10}{192}{5.2} & \tableRate{8}{192}{4.2} & 7 / 5 & \tablePval{3.9}{-1} & {\color{red}$\times$} \\
& & & Qwen3-32B & \tableRate{10}{192}{5.2} & \tableRate{5}{192}{2.6} & 6 / 1 & \tablePval{6.2}{-2} & {\color{red}$\times$} \\
& & & Qwen3.5-Flash & \tableRate{9}{192}{4.7} & \tableRate{3}{192}{1.6} & 6 / 1 & \tablePval{6.2}{-2} & {\color{red}$\times$} \\
& & & MiMo-V2.5 & \tableRate{10}{192}{5.2} & \tableRate{7}{192}{3.6} & 6 / 4 & \tablePval{3.8}{-1} & {\color{red}$\times$} \\
& & & GLM-4.7-Flash & \tableRate{1}{192}{0.5} & \tableRate{0}{192}{0.0} & 1 / 0 & \tablePval{5.0}{-1} & {\color{red}$\times$} \\
\noalign{\hrule height 1.5pt}
\end{tabular}
\end{table*}

\partitle{Trace-level detection} After a successful model-level detection, we further examine whether \name{} can identify the specific trajectories in the suspicious set $\mathcal{S}$ that are most likely used for distillation. We separately rank the scores $S_\tau$ of the 100 trajectories from $\mathcal{S}$ on the four student models and report the arithmetic mean of $\operatorname{\mathtt{Precision}}@K$ across the four students in Table~\ref{tab:trace-attribution}. The overall average precision across the six cohorts remains at least $82.1\%$ for every reported $K$. SWE-bench--GPT achieves a perfect $100.0\%$ at P@10. These results show that \name{} can identify the specific leaked data with high precision across all six cohorts.

\par\medskip
\noindent
\begin{minipage}{\textwidth}
\refstepcounter{table}
\label{tab:trace-attribution}
\noindent\textbf{Table~\thetable:} Trace-level attribution accuracy. Each entry is the arithmetic mean of $\operatorname{\mathtt{Precision}}@K$ over the four student models in the cohort.\par\smallskip
\small
\setlength{\tabcolsep}{5.5pt}
\begin{tabular*}{\linewidth}{@{\extracolsep{\fill}}l l | c c c c c@{}}
\noalign{\hrule height 1.5pt}
\textbf{Teacher Agent $\mathcal{M}_t$} & \textbf{Dataset} & \textbf{P@10} & \textbf{P@15} & \textbf{P@20} & \textbf{P@25} & \textbf{P@30} \\
\hline
\multirow{3}{*}{GPT-OSS-120B} & BFCL & 72.5\% & 80.0\% & 80.0\% & 79.0\% & 81.7\% \\
& SWE-bench & 100.0\% & 98.3\% & 96.2\% & 96.0\% & 94.2\% \\
& Telecom & 75.0\% & 78.3\% & 80.0\% & 82.0\% & 84.2\% \\
\hline
\multirow{3}{*}{Kimi-K2.5} & BFCL & 82.5\% & 81.7\% & 82.5\% & 84.0\% & 80.8\% \\
& SWE-bench & 85.0\% & 86.7\% & 86.2\% & 82.0\% & 80.8\% \\
& Telecom & 80.0\% & 83.3\% & 82.5\% & 75.0\% & 70.8\% \\
\hline
\multicolumn{2}{l|}{\textbf{Overall average}} & \textbf{82.5\%} & \textbf{84.7\%} & \textbf{84.6\%} & \textbf{83.0\%} & \textbf{82.1\%} \\
\noalign{\hrule height 1.5pt}
\end{tabular*}
\end{minipage}

\begin{figure}[htbp]
\centering
\begin{subfigure}[t]{0.32\textwidth}
\centering
\includegraphics[width=\linewidth]{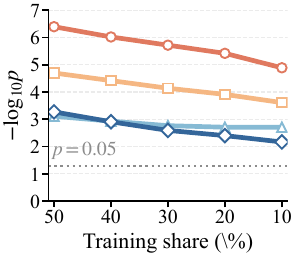}
\caption{BFCL / GPT}
\end{subfigure}\hfill
\begin{subfigure}[t]{0.32\textwidth}
\centering
\includegraphics[width=\linewidth]{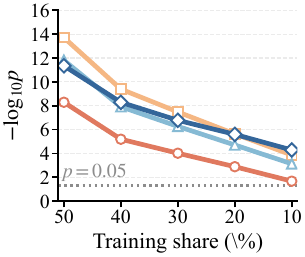}
\caption{SWE-bench / GPT-OSS-120B}
\end{subfigure}\hfill
\begin{subfigure}[t]{0.32\textwidth}
\centering
\includegraphics[width=\linewidth]{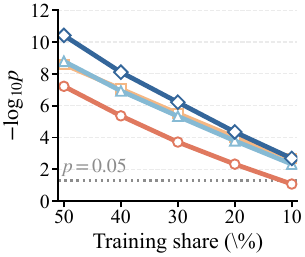}
\caption{Telecom / GPT}
\end{subfigure}

\begin{subfigure}[t]{0.32\textwidth}
\centering
\includegraphics[width=\linewidth]{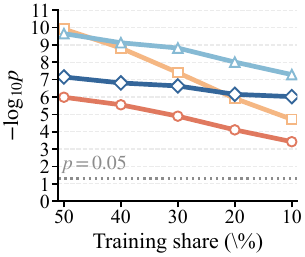}
\caption{BFCL / Kimi}
\end{subfigure}\hfill
\begin{subfigure}[t]{0.32\textwidth}
\centering
\includegraphics[width=\linewidth]{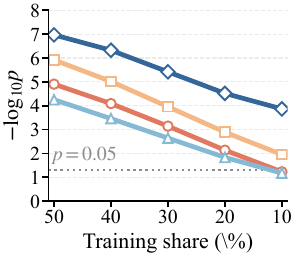}
\caption{SWE-bench / Kimi-K2.5}
\end{subfigure}\hfill
\begin{subfigure}[t]{0.32\textwidth}
\centering
\includegraphics[width=\linewidth]{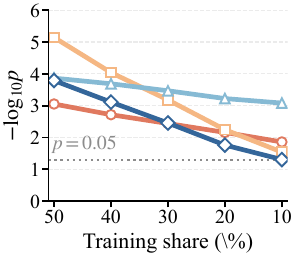}
\caption{Telecom / Kimi}
\end{subfigure}
\caption{
Impact of identification accuracy. The horizontal axis denotes the training share and the vertical axis denotes $-\log_{10}p$, where larger values indicate stronger evidence. In (a)--(f), the colored lines and hollow markers denote student models: \textcolor[HTML]{E0795F}{\rule[0.5ex]{0.5em}{1.5pt}\hspace{0.08em}$\circ$\hspace{0.08em}\rule[0.5ex]{0.5em}{1.5pt}} Mistral-24B, \textcolor[HTML]{F5B783}{\rule[0.5ex]{0.5em}{1.5pt}\hspace{0.08em}$\square$\hspace{0.08em}\rule[0.5ex]{0.5em}{1.5pt}} GLM-4.7-Flash, \textcolor[HTML]{86BAD4}{\rule[0.5ex]{0.5em}{1.5pt}\hspace{0.08em}$\triangle$\hspace{0.08em}\rule[0.5ex]{0.5em}{1.5pt}} Qwen3-14B, and \textcolor[HTML]{34669A}{\rule[0.5ex]{0.5em}{1.5pt}\hspace{0.08em}$\diamond$\hspace{0.08em}\rule[0.5ex]{0.5em}{1.5pt}} Qwen3-32B. The \textcolor[HTML]{8C8C8C}{\hdashrule[0.5ex]{1.2em}{1.2pt}{1pt 1.5pt}}  marks $p=0.05$.
}
\label{fig:training-share-sensitivity}
\end{figure}

\partitle{Impact on identification accuracy} To evaluate the impact of identification accuracy on detection accuracy, we simulate different training shares by resampling our empirical results. In practice, upstream traffic monitoring is often tuned for high recall to avoid missing distillation requests, which can introduce additional benign or unrelated trajectories into the suspicious set $\mathcal{S}$. We test \name{} under lower shares down to 10\%. Specifically, we divide the evaluated cards for each cohort and student model into trained and untrained groups. Next, we estimate their outcome distributions and resample these two groups at various target ratios. The $50\%$ share serves as the observed baseline. For other simulated shares, we perform 2,000 independent draws and apply the same one-sided exact sign test. As shown in Figure~\ref{fig:training-share-sensitivity}, which reports the median $p$-value, all 24 settings remain significant at a $20\%$ training share. At $10\%$, 21 of 24 settings remain significant.

\FloatBarrier

\subsection{Supplementary of Robustness}
\label{sec:supplementary-robustness}
\begin{figure}[htbp]
\centering
\begin{subfigure}[t]{\robustnessPanelWidth}
\centering
\includegraphics[width=0.9\linewidth]{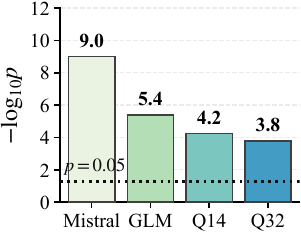}
\caption{BFCL: R1}
\end{subfigure}\hfill
\begin{subfigure}[t]{\robustnessPanelWidth}
\centering
\includegraphics[width=0.9\linewidth]{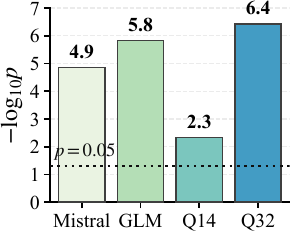}
\caption{BFCL: R2}
\end{subfigure}\hfill
\begin{subfigure}[t]{\robustnessPanelWidth}
\centering
\includegraphics[width=0.9\linewidth]{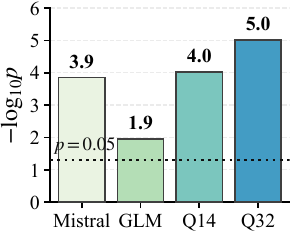}
\caption{BFCL: R3}
\end{subfigure}\par\vspace{-0.35em}
\begin{subfigure}[t]{\robustnessPanelWidth}
\centering
\includegraphics[width=0.9\linewidth]{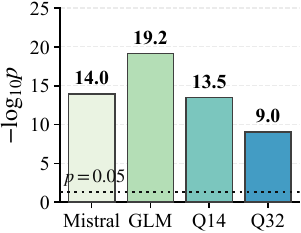}
\caption{SWE-bench: R1}
\end{subfigure}\hfill
\begin{subfigure}[t]{\robustnessPanelWidth}
\centering
\includegraphics[width=0.9\linewidth]{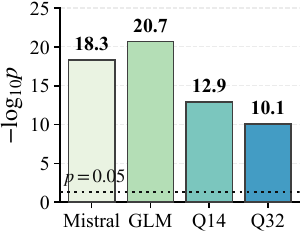}
\caption{SWE-bench: R2}
\end{subfigure}\hfill
\begin{subfigure}[t]{\robustnessPanelWidth}
\centering
\includegraphics[width=0.9\linewidth]{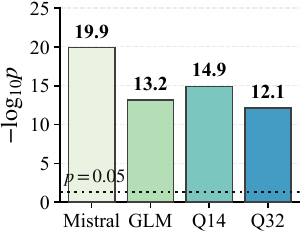}
\caption{SWE-bench: R3}
\end{subfigure}
\caption{Advanced data flooding robustness. The horizontal axis lists the four student models and the vertical axis reports $-\log_{10}p$. In (a)--(f), \robustnessModelLegend. The \textcolor[HTML]{8C8C8C}{\hdashrule[0.5ex]{1.2em}{1.2pt}{1pt 1.5pt}} marks $p=0.05$.}
\label{fig:data-flooding-plus-robustness}
\end{figure}

\partitle{Advanced data flooding} An attacker may build a diverse training set by collecting solutions to the same tasks from different teacher agents to improve the distilled model's performance~\cite{cui2023ultrafeedback, luo2026agentark}, which also dilutes the watermark signals. To evaluate the robustness of \name{} against this advanced data flooding tactic, we mix the original watermarked dataset $\mathcal{D}_c$ with clean trajectories generated by alternative teacher agents. Specifically, we sequentially introduce clean data from DeepSeek-V4-Flash, MiMo-V2.5, and Qwen3-235B-A22B. In the R1, R2, and R3 settings, we incorporate data from 1, 2, and 3 alternative teachers, resulting in ratios of $1{:}1$, $1{:}2$, and $1{:}3$. Then we use the combined dataset to train the student models and detect their watermarks. We evaluate a total of 24 settings across BFCL and SWE-bench, 3 dilution ratios, and 4 student models. As shown in Figure~\ref{fig:data-flooding-plus-robustness}, \name{} successfully detects every setting. Even under the most challenging R3 setting, all 8 tested student models remain successfully detected.

\begin{figure}[htp]
\centering
\begin{subfigure}[t]{\robustnessPanelWidth}
\centering
\includegraphics[width=0.9\linewidth]{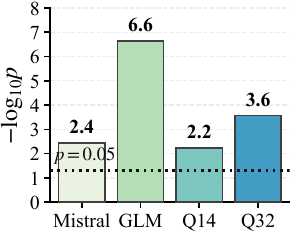}
\caption{BFCL: T10}
\end{subfigure}\hfill
\begin{subfigure}[t]{\robustnessPanelWidth}
\centering
\includegraphics[width=0.9\linewidth]{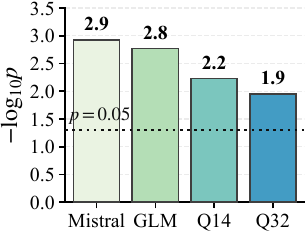}
\caption{BFCL: T15}
\end{subfigure}\hfill
\begin{subfigure}[t]{\robustnessPanelWidth}
\centering
\includegraphics[width=0.9\linewidth]{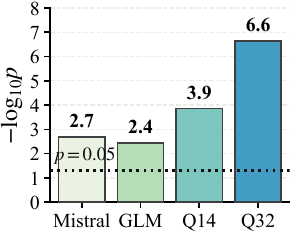}
\caption{BFCL: T20}
\end{subfigure}\par\vspace{-0.35em}
\begin{subfigure}[t]{\robustnessPanelWidth}
\centering
\includegraphics[width=0.9\linewidth]{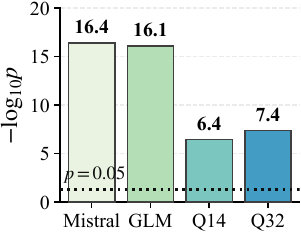}
\caption{SWE-bench: T10}
\end{subfigure}\hfill
\begin{subfigure}[t]{\robustnessPanelWidth}
\centering
\includegraphics[width=0.9\linewidth]{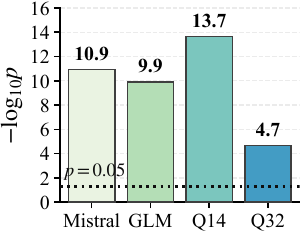}
\caption{SWE-bench: T15}
\end{subfigure}\hfill
\begin{subfigure}[t]{\robustnessPanelWidth}
\centering
\includegraphics[width=0.9\linewidth]{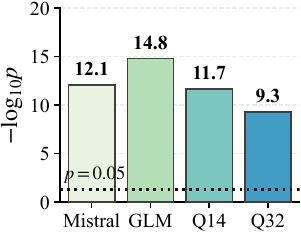}
\caption{SWE-bench: T20}
\end{subfigure}
\caption{Truncation robustness. The horizontal axis lists the four student models and the vertical axis reports $-\log_{10}p$ . In (a)--(f), \robustnessModelLegend. The \textcolor[HTML]{8C8C8C}{\hdashrule[0.5ex]{1.2em}{1.2pt}{1pt 1.5pt}} marks $p=0.05$.}
\label{fig:truncation-robustness}
\end{figure}

\partitle{Truncation} Before distillation, an attacker may truncate the trailing tokens of the interaction trajectories to fit the student models' context length limits or to deliberately disrupt potential watermark signals~\cite{kirchenbauer2023watermark}. To evaluate the robustness against this operation, we remove 10\%, 15\%, and 20\% of the tokens from the tail of each watermarked training sequence, and train the student models using these truncated trajectories. We evaluate a total of 24 settings across the BFCL and SWE-bench benchmarks, 3 truncation levels (T10, T15, and T20), and 4 student models. As shown in Figure~\ref{fig:truncation-robustness}, \name{} successfully detects the watermark in every setting. Even under the most aggressive T20 truncation setting, all 8 tested student models remain successfully detected.

\par\medskip
\begin{wraptable}{l}{0.50\textwidth}
\vspace{-\intextsep}
\refstepcounter{table}
\label{tab:baseline-robustness}
\noindent\textbf{Table~\thetable:} Agentwm robustness to data flooding. \par\smallskip
\centering
\scriptsize
\setlength{\tabcolsep}{4pt}
\begin{tabular*}{\linewidth}{@{\extracolsep{\fill}}l l | c c@{}}
\noalign{\hrule height 1.5pt}
\textbf{Dataset} & \textbf{Student Model $\mathcal{M}_s$} & \textbf{D1} & \textbf{D5} \\
\hline
\multirow{4}{*}{BFCL} & Mistral-24B & $0/5$ {\color{red}$\times$} & $0/5$ {\color{red}$\times$} \\
& GLM-4.7-Flash & $0/5$ {\color{red}$\times$} & $0/5$ {\color{red}$\times$} \\
& Qwen3-14B & $2/5$ {\color{red}$\times$} & $0/5$ {\color{red}$\times$} \\
& Qwen3-32B & $1/5$ {\color{red}$\times$} & $0/5$ {\color{red}$\times$} \\
\hline
\multirow{4}{*}{SWE-bench} & Mistral-24B & $1/5$ {\color{red}$\times$} & $1/5$ {\color{red}$\times$} \\
& GLM-4.7-Flash & $2/5$ {\color{red}$\times$} & $0/5$ {\color{red}$\times$} \\
& Qwen3-14B & $0/5$ {\color{red}$\times$} & $1/5$ {\color{red}$\times$} \\
& Qwen3-32B & $0/5$ {\color{red}$\times$} & $1/5$ {\color{red}$\times$} \\
\noalign{\hrule height 1.5pt}
\end{tabular*}
\vspace{-\intextsep}
\end{wraptable}
\partitle{Baseline robustness} Since Seqwm does not successfully detect distilled students(Table~\ref{tab:baseline-effectiveness}), we focus our robustness evaluation on Agentwm under data flooding (the D1 and D5 settings). To ensure a rigorous comparison, we process the clean standard data to match the initial distribution $P_c$ before mixing it with the watermarked traces. As shown in Table~\ref{tab:baseline-robustness}, all 16 evaluated settings fail to meet Agentwm's $\geq 3/5$ detection threshold. Agentwm encodes its watermark by shifting the tool-variant distribution towards $P_{\mathtt{wm}}$. By flooding the training corpus with clean trajectories calibrated to $P_c$, the overall frequencies are pulled back towards the unwatermarked state. This directly dilutes the artificial distributional signal, resulting in the distilled students not sufficiently reproducing the watermarked patterns for reliable detection.

\section{Discussion on Statistical Test}
\label{sec:statistical-test-discussion}

\par\medskip

\begin{wraptable}{l}{0.48\textwidth}
\vspace{-\intextsep}

\refstepcounter{table}
\label{tab:statistical-unit-sensitivity}

\noindent\textbf{Table~\thetable:} Statistical-unit sensitivity. Entries count settings with an unadjusted one-sided sign-test $p<0.05$; the 15\% training-share row uses the median over 2,000 resamples.\par\smallskip

\centering
\scriptsize
\setlength{\tabcolsep}{2.5pt}

\begin{tabular*}{\linewidth}{@{\extracolsep{\fill}}p{0.38\linewidth} c c c@{}}
\noalign{\hrule height 1.5pt}

\textbf{Condition} & \textbf{Card-level} & \textbf{Trace-level} & \textbf{Probe-level} \\

\hline

True positives & 24/24 & 24/24 & 24/24 \\
False positives & 0/48 & 0/48 & 3/48 \\
Paraphrasing attack & 8/8 & 7/8 & 8/8 \\
15\% training share & 24/24 & 22/24 & 24/24 \\

\noalign{\hrule height 1.5pt}
\end{tabular*}

\vspace{-\intextsep}
\end{wraptable}

To evaluate how the choice of statistical unit affects detection results, we perform the same paired-probe tests at three granularities: card, trace, and probe levels. The card-level test aggregates parameterized probes derived from the same card, the trace-level test further aggregates all cards from the same trajectory, and the probe-level test directly treats each paired probe as an individual statistical unit. As shown in Table~\ref{tab:statistical-unit-sensitivity}, the card- and trace-level tests remain consistent in the main effectiveness evaluation: both achieve 24/24 detections in the true-positive settings and 0/48 significant results in the false-positive settings. Differences mainly appear under more challenging conditions. Trace-level aggregation compresses multiple watermark releases from the same trajectory into a single sign, allowing evidence from different cards to cancel and thereby reducing statistical power and robustness. Consequently, detection decreases from 8/8 to 7/8 under the paraphrasing attack, and from 24/24 to 22/24 at a 15\% training share. In contrast, although probe-level testing shows stronger apparent statistical power, it repeatedly counts parameterized probes that share the same underlying release behavior and context, resulting in 3/48 cases with $p<0.05$ in the false-positive settings. Although none of these cases exceeds the final fullhit-gap threshold and therefore no actual false-positive decision is triggered, the results indicate a higher false-positive risk when probes are counted individually. Based on these observations, we use the card as the primary statistical unit: it avoids repeated counting of the same watermark release event while preserving effective evidence from distinct release events.

\FloatBarrier
\flushbottom

\clearpage
\section{Detailed Watermarking Protocols}

\begin{figure}[H]
\hrule width \hsize \kern 1mm \hrule width \hsize height 2pt 
\vspace{2mm}
\textbf{\name{} Watermark Embedding Protocol}
\vspace{1mm}
\hrule
\vspace{2mm}
\small
\textbf{Inputs:} Tool schema $\mathcal{F}$, Secret key $k$, Trajectory ID $\mathtt{id}_\tau$, Hyperparameters $(p_0, \Delta p, p_{\mathtt{max}}, B, K)$. \\
\textbf{Outputs:} Watermarked trajectory $\tau$, Private evidence card set $\mathcal{C}_\tau$.

\vspace{2mm}
\noindent \textbf{1. Initialization Phase:} 
\begin{itemize}
    \item \textit{System Setup:} 
    \begin{itemize}
        \item Evaluate the tool schema $\mathcal{F}$ using the safety model $\mathcal{M}_{\mathtt{safe}}$ to obtain the valid subset $\mathcal{F}_{\mathtt{safe}}$.
        \item Initialize the dynamic trigger probability $p_1 = p_0$, the remaining budget $b_1 = B$, and an empty evidence set $\mathcal{C}_\tau = \emptyset$.
    \end{itemize}
\end{itemize}

\noindent \textbf{2. Real-Time Embedding Phase:} \\
For each step $t$ immediately following a completed core action $a_{\mathtt{core}}^{t-1}$, the system executes:
\begin{itemize}
    \item \textit{Dynamic Scheduling:}
    \begin{itemize}
        \item Compute a cryptographic hash $u_t = \mathtt{Hash}(k \parallel a \parallel \mathtt{id}_\tau \parallel t) \in [0, 1)$.
        \item If the budget is none ($b_t = 0$) or the condition is not met ($u_t \ge p_t$), skip and failure update.
    \end{itemize}
    
    \item \textit{Candidate Generation \& Scoring:}
    \begin{itemize}
        \item Generate $K$ auxiliary candidates $(t_{\mathtt{aux}}^j, a_{\mathtt{aux}}^j)_{j=1}^K$ using the teacher model $\mathcal{M}_t$ and filter them through the validation mechanism.
        \item For each valid candidate $(t_{\mathtt{aux}}^j, a_{\mathtt{aux}}^j)$, compute the score $S_j$ based on naturalness ($s_{\mathtt{rel}}^j$), logical consistency ($s_{\mathtt{relb}}^j$), and repeatability $s_{\mathtt{relb}}^j$ for the tool pairing $(f_{\mathtt{core}}, f_{\mathtt{aux}}^j)$.
        \item Select the candidate with the highest score as $(t_{\mathtt{aux}}^{\mathtt{best}}, a_{\mathtt{aux}}^{\mathtt{best}})$. If no valid candidates exist, skip to the failure update.
    \end{itemize}

    \item \textit{Injection \& Success Update:}
    \begin{itemize}
        \item Inject $(t_{\mathtt{aux}}^{\mathtt{best}}, a_{\mathtt{aux}}^{\mathtt{best}})$ into the trajectory to receive the auxiliary observation $o_{\mathtt{aux}}^{\mathtt{best}}$.
        \item Reset the probability $p_{t+1} = p_0$ and decrement the budget $b_{t+1} = b_t - 1$.
        \item Increment the historical repeat count for this specific tool pairing $(f_{\mathtt{core}}, f_{\mathtt{aux}}^{\mathtt{best}})$.
        \item Create an evidence card $c_t = (h, t_{\mathtt{core}}^{t-1}, a_{\mathtt{core}}^{t-1}, o_{\mathtt{core}}^{t-1}, f_{\mathtt{aux}}^{\mathtt{best}}, p_{\mathtt{aux}}^{\mathtt{best}})$ and append it to $\mathcal{C}_\tau$.
    \end{itemize}

    \item \textit{Failure/Skip Update (Executed only if no injection occurred):}
    \begin{itemize}
        \item Increase the trigger probability $p_{t+1} = \min\{p_t + \Delta p, p_{\mathtt{max}}\}$ to boost future injection chances, and retain the current budget $b_{t+1} = b_t$.
    \end{itemize}
\end{itemize}
\vspace{1mm}
\hrule width \hsize \kern 1mm \hrule width \hsize height 2pt 
\caption{The watermark embedding protocol of \name{}. }
\label{fig:watermark_embedding_protocol}
\end{figure}

\clearpage
\begin{figure}[H]
\hrule width \hsize \kern 1mm \hrule width \hsize height 2pt 
\vspace{2mm}
\textbf{\name{} Watermark Detection Protocol}
\vspace{1mm}
\hrule
\vspace{2mm}
\small
\textbf{Inputs:} Combined evidence pool $\mathcal{C}_{\mathcal{S}}$, Suspect model $\mathcal{M}_s$, Teacher model $\mathcal{M}_t$, Hyperparameters $(p, r, N)$. \\
\textbf{Outputs:} Detection decision $\in \{\mathtt{True}, \mathtt{False}\}$, and the statistical $p\mathtt{-value}$.

\vspace{2mm}
\noindent \textbf{1. Probe Construction Phase:} \\
For each evidence card $c_i \in \mathcal{C}_{\mathcal{S}}$, the system executes:
\begin{itemize}
    \item \textit{Paired Probes Generation:}
    \begin{itemize}
        \item Generate up to $N_i$ parameterized variants using card $c_i$ to form the real probe set $p_{\mathtt{real}}^i$.
        \item Construct the corresponding fake probe set $p_{\mathtt{fake}}^i$ by replacing the core action in $p_{\mathtt{real}}^i$ with a semantically altered fake core.
    \end{itemize}
\end{itemize}

\noindent \textbf{2. Evaluation Phase:} \\
For each probe variant $j \in [1, N_i]$ of card $c_i$, the system executes:
\begin{itemize}
    \item \textit{Target Model Query \& Parsing:}
    \begin{itemize}
        \item Submit the real probe $p_{\mathtt{real}, j}^i$ to the suspect model $\mathcal{M}_s$ and parse the output to compute the strict fullhit indicator $h_{i,j}^{\mathtt{real}} \in \{0, 1\}$.
        \item Submit the fake probe $p_{\mathtt{fake}, j}^i$ to the suspect model $\mathcal{M}_s$ and parse the output to compute the strict fullhit indicator $h_{i,j}^{\mathtt{fake}} \in \{0, 1\}$.
    \end{itemize}
\end{itemize}

\noindent \textbf{3. Card-Level Aggregation Phase:} \\
For each card $c_i \in \mathcal{C}_{\mathcal{S}}$, the system executes:
\begin{itemize}
    \item \textit{Hit Rate \& Performance Calculation:}
    \begin{itemize}
        \item Compute the hit rates across all $N_i$ variants: $R_i^{\mathtt{real}} = \frac{1}{N_i} \sum_{j} h_{i,j}^{\mathtt{real}}$ and $R_i^{\mathtt{fake}} = \frac{1}{N_i} \sum_{j} h_{i,j}^{\mathtt{fake}}$.
        \item Calculate the performance difference: $\Delta_i = R_i^{\mathtt{real}} - R_i^{\mathtt{fake}}$.
        \item Tally the results: increment the real win count $w$ if $\Delta_i > 0$, and increment the fake win count $\ell$ if $\Delta_i < 0$ (ties where $\Delta_i = 0$ are excluded).
    \end{itemize}
\end{itemize}

\noindent \textbf{4. Statistical Test \& Decision Phase:} 
\begin{itemize}
    \item \textit{Hypothesis Testing:}
    \begin{itemize}
        \item Compute the one-sided exact sign-test $p\mathtt{-value} = \sum_{k=w}^{w+\ell}\binom{w+\ell}{k}2^{-(w+\ell)}$.
        \item Compute the global mean difference $\Delta_{\mathtt{mean}} = \frac{1}{|\mathcal{C}_{\mathcal{S}}|} \sum_{i} \Delta_i$.
    \end{itemize}
    
    \item \textit{Dual Criteria Check:}
    \begin{itemize}
        \item Output $\mathtt{True}$ (Detection Successful) if both $p\mathtt{-value} < p$ and $\Delta_{\mathtt{mean}} \ge r$ hold. Otherwise, output $\mathtt{False}$.
    \end{itemize}
\end{itemize}
\vspace{1mm}
\hrule width \hsize \kern 1mm \hrule width \hsize height 2pt 
\caption{The watermark detection protocol of \name{}.}
\label{fig:watermark_detection_protocol}
\end{figure}

\clearpage

\section{Limitations of Unwatermarked Distillation Detection}
\label{sec:similarity-attribution}

We additionally investigate approaches that infer distillation from
similarity to a suspected teacher without an embedded watermark.
We examine two representative similarity-based analyses of distillation:
token-level $n$-gram overlap under reasoning prefilling
\citep{panfilov2026stealing} and agent execution graph similarity
\citep{yang2026agents}. These approaches assess similarity at the levels of
surface text and structured tool-use behavior, respectively.
Our analyses show that shared solution patterns and response formatting
can yield high textual overlap, while structurally different dependency
graphs can receive near-perfect similarity scores. High scores therefore
admit multiple explanations and do not uniquely identify a training source.

\subsection{Vulnerabilities in Token-Level N-gram Matching}
\label{sec:textual-similarity}

\citet{panfilov2026stealing} study visible-answer overlap after prefilling a
model's reasoning channel with a short teacher reasoning prefix. Their
Appendix~B.2 evaluates 30 Humanity's Last Exam problems using shared
1-, 2-, and 3-grams against the first 100 tokens of the reference answer,
with best-of-$k$ sampling up to $k=100$. The visible answer is generated
entirely by the target model. Their Figures~11--23 illustrate 13 public
examples with problem statements, teacher reasoning excerpts, and
visible-answer excerpts. Using selected public cases, we test whether
alternative prefixes and shared solution patterns can produce high
overlap with the same teacher reference.

We use a fixed local overlap implementation. For each candidate $a$ and
reference $b$, we tokenize both texts with the same regex tokenizer and
retain up to their first 100 tokens. Let $G_n(x)$ denote the set of distinct
$n$-grams in the retained tokens of $x$. We compute
\begin{equation}
    s_{\mathtt{ngram}}(a,b)
    =
    \frac{\sum_{n=1}^{3}|G_n(a)\cap G_n(b)|}
         {\sum_{n=1}^{3}|G_n(b)|}.
    \label{eq:local-ngram-proxy}
\end{equation}

\begin{figure}[H]
    \centering
    \includegraphics[width=\textwidth]{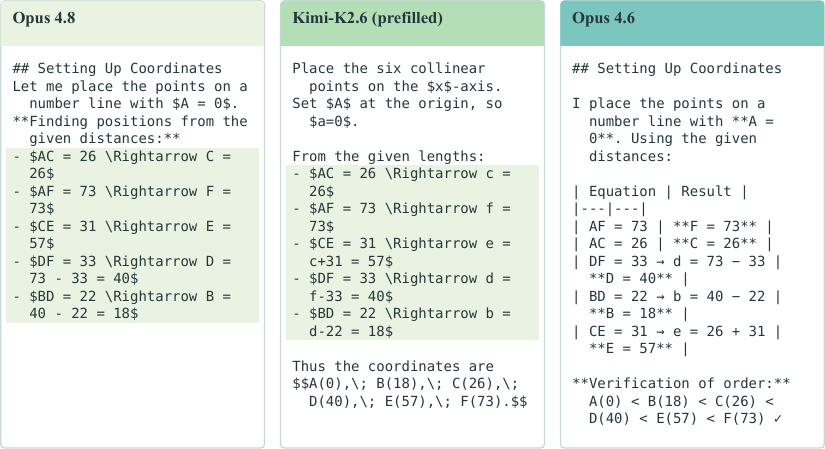}
    \caption{Excerpts from the original answers to the geometry problem in
    Figure~11 of \citet{panfilov2026stealing}.
    Green shading marks the shared $C,F,E,D,B$ derivation order.
    For each candidate model, we show the highest-overlap answer from four
    samples.}
    \label{fig:geometry-answer-comparison}

    \vspace{10pt}
    \includegraphics[width=\textwidth]{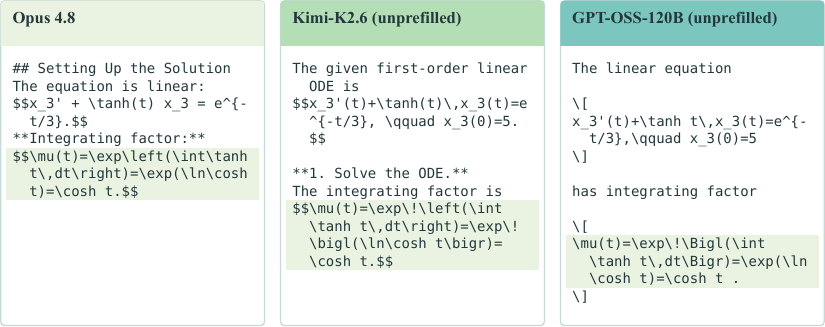}
    \caption{Excerpts from the original answers to the differential-equation
    problem in Figure~13 of \citet{panfilov2026stealing}.
    Green shading marks the shared $\cosh t$ integrating factor.
    Each candidate has the highest reported overlap among four unprefilled
    samples.}
    \label{fig:ode-answer-comparison}
\end{figure}

\partitle{Alternative prefixes and comparisons across model versions}
On the geometry problem in Figure~11 of \citet{panfilov2026stealing},
Kimi-K2.6, which was publicly released before Claude Opus~4.8
\citep{team2025kimi,anthropic2026opus48},
obtains mean/best-of-4 scores of $0.274/0.565$ with the
teacher opening \texttt{This is a known}, compared with $0.373/0.662$
under \texttt{This is an unfamiliar} and $0.250/0.415$ without a prefix.
The best local score of $0.662$ approaches the reported $0.80$ overlap
between prefilled Kimi-K3 and Opus~4.8.
Within this fixed scoring setup, the alternative prefix produces a larger
increase than the authentic prefix across both metrics.
The selected $0.662$ answer shares the opening derivation order $C,F,E,D,B$
with the Opus~4.8 reference, as shown in Figure~\ref{fig:geometry-answer-comparison}.
The sampled Opus~4.6 answers use tables or
different orders; the highest score among these answers is $0.343$ against the
same reference. Thus, this local formatting match is not exclusive to the
teacher prefix or to the sampled Claude-family answers.

\partitle{Convergence on standard solutions}
For the linear differential equation in Figure~13 of
\citet{panfilov2026stealing}, a standard solution uses a $\cosh$ integrating
factor. Without a teacher prefix,
GPT-OSS-120B reaches a mean overlap of $0.500$ and a best-of-4 score of
$0.537$. Kimi-K2.6 similarly scores $0.501$ without a prefix and $0.502$
with the teacher prefix. The unprefilled local scores are numerically close
to the reported $0.50$ overlap between prefilled Kimi-K3 and Opus~4.8.
The unprefilled answer excerpts are shown in Figure~\ref{fig:ode-answer-comparison}.
Our results show that substantial local overlap can arise in standard
solutions with conventional terminology and ordering, while the teacher
prefix adds little in the Kimi-K2.6 comparison.

\partitle{Response steering and sampling}
Visible-answer overlap must also be interpreted in light of how responses
are generated, presented, and selected. In our geometry evaluation of Kimi-K3,
all 20 completions across five prefix
conditions re-derive the solution, and none states the answer within its
first 50 reasoning tokens, unlike the early answer in the reference trace.
Higher visible overlap can therefore coexist with a different observed
reasoning pattern. At the level of answer presentation, even correct
solutions can receive very low scores when little matching text is
available: for the problem in Figure~15 of \citet{panfilov2026stealing},
four correct answers constrained to at most
15 words each score $0.005$. Such sensitivity to presentation also matters
when interpreting best-of-$k$ results, which report the largest overlap
in a sampling pool. Formatting cues can therefore raise the reported score
through occasional high-overlap completions without representing typical
output behavior.
\subsection{Vulnerabilities in Agent Execution Graph Similarity}
\label{sec:dependency-ambiguity}
AgentEcho's Action Graph Similarity (AGS) averages optional-tool agreement
$S_{\mathtt{node}}$, sequential-pattern similarity $S_{\mathtt{seq}}$,
and dependency-pattern similarity $S_{\mathtt{dep}}$
\citep{yang2026agents}. Moving from text to tool behavior introduces
structural evidence, but its interpretation depends on which graph
properties the metrics preserve.

\partitle{Scale invariance and structural ambiguity}
For a dependency graph $G$, $S_{\mathtt{dep}}$ compares
$\phi(G)=(r_{\mathtt{reuse}},d_{\mathtt{max}},r_{\mathtt{fanout}})$ by cosine
similarity \citep[Appendix~B.4]{yang2026agents}. These features describe
the fraction of calls after the first that receive dependency inputs,
the longest dependency path, and the fraction of
source nodes with multiple outgoing dependency edges. Cosine similarity
ignores feature-vector magnitude. Consider two graphs with 11 nodes:
one contains a single dependency edge, whereas the other is a complete
10-edge chain. With path length measured in edges,
\begin{equation}
    \phi(G_{\mathtt{single}})=(0.1,1,0),\qquad
    \phi(G_{\mathtt{chain}})=(1,10,0)
    =10\phi(G_{\mathtt{single}}).
    \label{eq:dependency-counterexample}
\end{equation}
Their $S_{\mathtt{dep}}$ is exactly 1 despite large differences in reuse
and dependency depth. Hence, a perfect score does
not guarantee closely matching dependency structures.

\partitle{Evidence from tool trajectories}
We audit 75 trajectories from five models on 15 tasks and compare all
ten model pairs within each task, yielding 150 comparisons.
The mean $S_{\mathtt{dep}}$ is $0.885$,
and 121 comparisons score at least $0.9$. Among these, 53 have dependency
edge counts differing by at least three, and 49 pair a successful
trajectory with a failed one. In \texttt{retail-37}, a failed trajectory
with one dependency edge and a successful trajectory with 12 edges score
$0.991$. We further verify candidate dependency edges with an LLM judge,
using anonymized values and a prompt adapted from the published protocol
\citep[Appendix~B.1]{yang2026agents}.
Both the score and the edge-count discrepancy remain unchanged for this
example.
Thus, near-perfect dependency similarity persists despite substantial
structural differences, even after LLM-based edge verification.

\partitle{Behavioral convergence after reasoning distillation}
\label{sec:reasoning-distillation-convergence}
Inspired by the reference-based formulation of \citet{rawat2026reference},
we measure changes in behavioral similarity relative to the base checkpoint.
We use the Qwen3.5-9B base model as the reference checkpoint and evaluate
its publicly released reasoning-distilled variant.\footnote{Model card:
\href{https://huggingface.co/Jackrong/Qwen3.5-9B-Claude-4.6-Opus-Reasoning-Distilled-v2-GGUF}{Jackrong's
reasoning-distilled Qwen3.5-9B (GGUF)}.}
The model card identifies Claude Opus~4.6 as the teacher.
Our audit of 12,592 publicly available examples from the model card's
listed datasets finds no structured tool trajectories.
We use the same agent harness
and user simulator for 15 $\tau$-bench-style tasks, with one rollout per
model and task. Claude Sonnet~4.5 (thinking), Kimi-K2 (thinking), and
GPT-OSS-120B serve as comparison models.

\begin{table}[htbp]
    \centering
    \small
    \setlength{\tabcolsep}{5pt}
    \begin{tabular}{lrrrrrr}
        \hline
        Comparison model & AGS$_{\mathtt{base}}$ & AGS$_{\mathtt{dist}}$
        & $\Delta\mathtt{AGS}$ & $\Delta S_{\mathtt{node}}$
        & $\Delta S_{\mathtt{seq}}$ & $\Delta S_{\mathtt{dep}}$ \\
        \hline
        Claude Sonnet~4.5 (thinking) & 0.549 & 0.709 & +0.160 & $-0.031$ & +0.425 & +0.086 \\
        Kimi-K2 (thinking) & 0.561 & 0.850 & +0.289 & +0.042 & +0.750 & +0.077 \\
        GPT-OSS-120B & 0.584 & 0.770 & +0.186 & $-0.047$ & +0.525 & +0.080 \\
        \hline
    \end{tabular}
    \caption{Changes in AGS and its components after reasoning
    distillation on the same 15 tasks.
    Component values are rounded independently.}
    \label{tab:reasoning-distillation-similarity}
\end{table}

Table~\ref{tab:reasoning-distillation-similarity} shows that AGS increases
toward all three models, alongside a rise in task success from
$4/15$ to $12/15$. The gains are dominated by $S_{\mathtt{seq}}$, which
increases by $0.425$--$0.750$.
In contrast, $S_{\mathtt{dep}}$ increases by $0.077$--$0.086$, while
$S_{\mathtt{node}}$ decreases toward both Claude and GPT-OSS.
Thus, these gains primarily reflect convergence in local execution
statistics and do not uniquely identify teacher-specific tool-use inheritance.

\partitle{Implications for attribution}
These cases expose ambiguity in interpreting high unwatermarked similarity
as evidence of a specific training source. They motivate controls for shared
solutions, response presentation, and general behavioral improvement,
alongside calibrated false-positive rates. Our watermarking framework
provides private verification signals tied to recorded teacher interactions.

\end{document}